\documentclass[12pt,hidelinks]{article}
\usepackage{amsmath, amsthm, amssymb}
\usepackage{graphicx,psfrag,epsf}
\usepackage{enumerate}
\usepackage{natbib}
\usepackage{bm}
\usepackage{bbm}
\usepackage{url} 
\usepackage{xcolor}
\usepackage{tikz}
\usepackage{booktabs, multirow, threeparttable,easyReview,ragged2e, float}
\usepackage{tkz-graph}
\usepackage{stmaryrd}
\usepackage{subfig}
\usepackage{caption}
\usepackage{bbm}
\usepackage[colorlinks=true]{hyperref}					
\hypersetup{%
        urlcolor=blue,
	linkcolor=black,%
	citecolor=black}
\usepackage{url} 
\usepackage[symbol*, bottom]{footmisc}
\usepackage{appendix}



\usepackage{etoolbox}
\AtBeginEnvironment{appendices}{%
     
    \makeatletter

    \makeatother

    \setcounter{section}{0}  
    \setcounter{equation}{0}
    \setcounter{theorem}{0}
    \setcounter{lemma}{0}
    \setcounter{assumption}{0}
    \setcounter{definition}{0}
}

\DefineFNsymbolsTM{myfnsymbols}{
  \textdagger    \dagger
  \textdaggerdbl \ddagger
  \textsection   \mathsection
   \textasteriskcentered *
}%
\setfnsymbol{myfnsymbols}

\newcommand{\blind}{1}

\def\pr{\operatorname{pr}}

\theoremstyle{plain}
\newtheorem{assumption}{\bf Assumption}

\newtheorem{example}{\bf Example}
\newtheorem{theorem}{\bf Theorem}
\newtheorem{definition}{\bf Definition}

\begin{document}

\def\spacingset#1{\renewcommand{\baselinestretch}%
{#1}\small\normalsize} \spacingset{1}


\if1\blind
{
  \title{\bf Causal inference with dyadic data in randomized experiments\\
  \vspace{1pt}}
  \author{Yilin Li$^{1,2}$, Lu Deng$^{2}$, Yong Wang$^{2}$, and Wang Miao$^{1}$\footnote{Address for correspondence: Wang Miao, Department of Probability and Statistics, Peking University, 5 Yiheyuan Road, Beijing 100871, P.R. China. Email: \href{mwfy@pku.edu.cn}{mwfy@pku.edu.cn}}
    \vspace{10pt}
\\
{ Department of Probability and Statistics, Peking University$^1$}\\
{ Tencent Inc.$^2$}}
  \maketitle
} \fi

\spacingset{1.3} 

\begin{abstract}
Estimating treatment effects in networked settings is a central challenge in online controlled experiments, particularly on social media platforms. We investigate a scenario where the unit-level outcome of interest comprises a series of dyadic outcomes that record pairwise interactions between units, spanning from point-to-point messaging at the microscale to bilateral trade flows at the macroscale. Because the response is defined at the dyadic level, the treatment assigned to one unit can affect the outcomes of all dyads that involve it, inducing a form of network interference. We propose a design-based causal inference framework for randomized experiments with dyadic outcomes. Within this framework, we propose estimators of the global average treatment effect under Bernoulli, complete, and cluster randomization, derive the convergence rates, and establish a central limit theorem for Bernoulli randomization. We further construct a class of variance estimators that are asymptotically conservative under transparent degree conditions. Numerical studies show that the proposed estimators can reduce bias and mean squared error relative to estimators based on unit-level outcomes in a range of finite-sample settings. We illustrate the methods using two large-scale experiments on WeChat, evaluating the impact of a recommendation algorithm and a calling feature.
\end{abstract}

\noindent%

\section{Introduction}

Randomized experiments are widely used in online A/B testing and other digital experimentation settings, where they help accurately assess the impact of various interventions on basic functional tests, user experiences, and other facets of digital platforms \citep{gupta2019top,kohavi2020trustworthy}. Dozens of such experiments are conducted every day on major platforms, and one of the key characteristics of those that take place on social networks is interference: the treatment assigned to one user can affect the outcomes observed for other users. Interference complicates the representation of potential outcomes because the stable unit treatment value assumption  \citep[SUTVA,][]{rubin1980sutva} no longer holds, and it invalidates estimators that rely on SUTVA.  It is crucial in problems involving social networks and is prevalent in various applications, such as communication and sharing behaviour among individuals mediated by a complex social network, and the spread of an infectious disease through a network of human interactions \citep{ogburn2017vaccines}. It also arises naturally whenever the intervention operates through pairwise interactions, as in content sharing, collaborative tools, Voice over Internet Protocol (VoIP) services, and matching games that require mutual access \citep{karrer2021network,weng2024experimental}. In these scenarios, traditional unit-level randomization and corresponding estimators risk yielding biased feature value assessments due to interference effects.

Researchers have made significant progress in causal inference with interference, including various types of causal estimands \citep{hudgens2008interference, vanderweele2011effect, savje2021average, hu2022average} , models for interference mechanisms, and methods for identification and estimation \citep{eckles2016design, aronow2017estimating, liu2016inverse, bhattacharya2020causal,ogburn2020causal, bhattacharya2020causal, tchetgen2021auto, leung2022causal, li2022random, ogburn2024causal}. A commonly-used device is the exposure mapping, which summarizes the interference experienced by each unit through a low-dimensional function of the assignment vector \citep{manski2013socialinteraction, aronow2017estimating, chin2019regression}. However, specifying the exposure-mapping typically requires strong structural knowledge about how interference operates. \citet{savje2024causal} quantified the impact of misspecifying such a mapping. \citet{shpitser2024modeling} develop a symmetric treatment decomposition framework for separating spillover effects into direct and indirect components under interference, whereas our work focuses on design-based estimation and inference for causal effects with dyadic outcomes in randomized experiments. Complementary work addresses interference at the design stage, including cluster randomization \citep{ugander2013graph, ugander2023randomized}, ego-cluster experiments \citep{saintjacques}, and staggered roll-out designs \citep{cortez2022staggered, han2023detect}. In addition, there are other types of interference beyond network interference, such as carryover effects \citep{bojinov2019time,han2024population} and peer effects \citep{goldsmith2013social, egami2024identification, luo2025identification}. 
This growing body of research highlights the importance of addressing interference for reliable causal inference. 

The previous proposals for handling interference in social networks have mainly focused on the analysis of unit-level data.  In many social-network applications, however, the outcomes capturing interactions and relationships between pairs of units are naturally defined at the {dyadic} level, offering new opportunities for causal inference with interference. For example, the total number of messages received by a user can be disaggregated into the message counts exchanged between that user and each of their contacts. Dyad-level information preserves pair-specific variation and reveals complex interactions among units, thereby supporting more accurate policy decisions. Beyond social networks, the analysis of dyadic data is also appealing in various scientific fields, including cross-border equity flows \citep{portes2005determinants}, international trade relations \citep{carlson2024}, behavioural ecology \citep{apicella2012social}, and public health \citep{luke2007network}.

Dyadic data analysis has a long tradition in econometrics and statistics, primarily focusing on dependence-aware estimation and inference for regression and correlation models. \citet{fafchamps2007risk} studied identification issues in dyadic regression and proposed a dyadic-robust variance estimator that extends the spatial heteroskedasticity and autocorrelation consistent (HAC) estimator; \citet{cameron2014robust} applied it to international trade flows. \citet{aronow2015cluster} proposed a sandwich-type robust variance estimator for linear regression that accommodates the complex clustering structure of dyadic data, and \citet{tabord2019inference} established asymptotic properties of dyadic-robust $t$-statistics under more general conditions and asymptotic regimes. \citet{canen2024inference} proposed a linear dyadic regression model that explicitly accounts for dependence across indirectly linked dyads. For nonparametric dyadic regression, \citet{graham2021minimax} derived a minimax lower bound for the mean regression function and provided a dyadic Nadaraya--Watson estimator, and a series of recent works \citep{graham2024kernel,chiang2023empirical,cattaneo2024uniform}  studied kernel density estimation for dyadic data. {\citet{graham2020dyadic} provides a model-based treatment of dyadic outcomes that posits independent and identically distributed latent unit-level and dyad-level variables, with a brief discussion of interference in his Section~5.5. These contributions concern estimation and inference within dyadic regression or correlation models, rather than engaging with the causal aspects of dyadic data. Our framework differs in that it is design-based, treats the potential outcomes as fixed, and derives inference solely from the randomization mechanism.}  

In contrast to the vast literature on causal inference with unit-level outcomes and dyadic data analysis, the research on causal inference with dyadic data is relatively sparse.  \citet{DAmour2019causal} considered dyadic outcomes when treatment is assigned at the dyadic level, while we focus on the more common setting in which treatment is assigned at the unit-level. \citet{bajari2023} considered dyadic outcomes in a bipartite marketplace, where dyads connect two distinct populations (e.g., buyers and sellers) and proposed a multiple-randomization design tailored to that structure; we instead consider dyads within a single population, such as user-to-user pairs. \citet{weng2024experimental} studied experimental design problems on one-sided matching platforms by constructing a stochastic market model.  \citet{shi2024asymptotic} established a central limit theorem for the quadratic assignment procedure and developed associated tests for correlations of dyadic outcomes.

In this paper, we focus on the global average treatment effect in randomized experiments with dyadic outcomes. We develop a design-based causal inference framework for randomized experiments with dyadic data, without invoking the linear model. Our primary contributions include introducing the dyadic interference problems, formalizing the inference framework for dyadic data, and providing estimators with large-sample properties.  Section~\ref{sec:dyadinterfer} formalizes dyadic interference, defines the target estimand, and shows that conventional estimators based on aggregated unit-level outcomes are generally biased. Section~\ref{sec:estimator} introduces Horvitz--Thompson and H\'{a}jek estimators based directly on dyadic data and establishes their convergence rates under Bernoulli and complete randomization, together with asymptotic normality under Bernoulli randomization. In Section~\ref{sec:confidence}, we discuss the confidence statement for the estimators proposed under Bernoulli randomization and complete randomization, including a class of variance estimators. Sections~\ref{sec:simu} and~\ref{sec:real} evaluate the methods through numerical simulations and two large-scale WeChat experiments, one on a new recommendation algorithm and the other on a VoIP feature.

\section{Dyadic interference}\label{sec:dyadinterfer}

In some situations, the unit-level outcomes can be disaggregated into interactions between units. For example, the total volume of messages received by a user can be decomposed into message counts from individual contacts. Similarly, a user’s total call duration within a given week can be disaggregated into the contributions of each connected peer. To formalize such settings, consider a finite population of $n$ units. For each ordered pair $(i,j)$, let $Y_{ij}$ denote a dyadic outcome measuring the interaction from unit $i$ to unit $j$. The dyadic outcome may be a binary, count-valued, or continuous-valued variable recording the observed interactions. Depending on the application, interactions can be either directed, so that $Y_{ij}\neq Y_{ji}$, or undirected, so that $Y_{ij}=Y_{ji}$. Examples include message exchanges in social networks and symmetric measures of VoIP communication duration, respectively. Although unit-specific outcomes $Y_{ii}$ can also be accommodated, we restrict attention to dyadic outcomes with $i\neq j$ in the main development and revisit the setting with $Y_{ii}$ in Section~\ref{sec:dis}.
{We reserve an auxiliary symbol $\dagger$ to denote the absence of an interaction. Thus, $Y_{ij}=\dagger$ when units $i$ and $j$ do not interact, whereas $Y_{ij}$ takes a real value whenever an interaction is observed. For non-negative outcomes, such as message counts, call durations, or binary click indicators, one may equivalently encode the absence of an interaction by taking $\dagger=0$. When zero is itself a substantively meaningful observed value, $\dagger$ is kept as a distinct symbol in the aggregations below. The theoretical results are identical under either encoding.} Based on this, we define the upstream and downstream neighbour sets of unit $i$ according to the direction of the dyadic outcomes: $\mathcal{U}_{i} = \{j: Y_{ji} \neq \dagger\}$ and $\mathcal{D}_i = \{j: Y_{ij} \neq \dagger\}$, respectively. The total neighbour set of unit $i$ is then given by $\mathcal{N}_{i}=\mathcal{U}_{i}\cup\mathcal{D}_{i}$.

Many unit-level outcomes of practical interest are obtained by aggregating dyadic outcomes across neighbours. This motivates our focus on inference with additive dyadic outcomes. 
For additive dyadic data, we consider three unit-level aggregates:
\begin{equation}\label{eq:outcomes}
U_{i} = \sum_{j\in \mathcal{U}_{i}} Y_{ji}, \quad D_{i} = \sum_{j\in \mathcal{D}_{i}} Y_{ij},\quad Y_i = \sum_{j\in \mathcal{N}_{i}} (Y_{ij}+Y_{ji}) = U_{i} + D_{i}.
\end{equation}

\begin{example}\label{ex:1}
    In a content-sharing platform, the dyadic outcome $Y_{ij}$ counts links shared by user $i$ and clicked by user $j$. The upstream aggregate $U_{i}$ counts links received and clicked by user $i$, while the downstream aggregate $D_{i}$ counts links shared by user $i$ and clicked by other users.
\end{example}

\begin{example}\label{ex:2}
 For a VoIP service with pairwise call durations $Y_{ij}$, the aggregate $U_i$ is the duration of calls received by user $i$, $D_i$ is the duration of calls made by user $i$, and $Y_i$ is the total call duration involving user $i$. 
\end{example}
Some non-additive dyadic quantities can also be accommodated after a suitable transformation; for example, click-through probabilities may be analysed on a log-survival scale when an additive representation is appropriate.
\begin{example}[Continuation of Example~\ref{ex:1}]
Consider the probability that user $i$ clicks at least one shared link. Let ${\pr}_{ij} \in (0,1)$ be the predicted click-through probability for user $j$ on links from $i$. Assuming independence across links, this probability equals $1 - \prod_{j\in\mathcal{U}_i}(1-\pr_{ji})$. 
Define $Y_{ij} = \log(1-\pr_{ij})$. The unit-level outcome $U_i$ represents the log-likelihood of user $i$ not clicking any shared content.
\end{example}

Dyadic outcomes define a directed graph: units are nodes, and a directed edge $(i,j)$ exists if and only if $Y_{ij}\neq\dagger$. Figure~\ref{fig:example} illustrates the relationship between dyadic and aggregated unit-level outcomes.

\begin{figure}[H]
    \centering
\begin{tikzpicture}[scale=1,
->,
shorten >=1pt]
\node[circle, draw, rounded corners=0pt, fill=lightgray!41] (0) at(0,0){$\mathsf{1}$};
\node[circle, draw, rounded corners=0pt, fill=lightgray!41] (1) at(-2.,0){$\mathsf{2}$};
\node[circle, draw, rounded corners=0pt, fill=lightgray!41] (2) at(0,2){$\mathsf{3}$};
\node[circle, draw, rounded corners=0pt, fill=lightgray!41] (3) at(2.,0){$\mathsf{4}$};
\node (4) at(1.2,0.5){$Y_{14}$};
\node (4) at(-1.2,0.5){$Y_{21}$};
\node (4) at(0.7,1){$Y_{13}$};
\node (4) at(-0.7,1){$Y_{31}$};
\node (5) at(-1.65,1.6){$U_{1}$};
\node (6) at(1.65,1.6){$D_{1}$};
\draw[-{>[scale=2]}] (1) --(0);
\draw[-{>[scale=2]}] (2) --(0);
\draw[-{>[scale=5]}] (0) --(2);
\draw[-{>[scale=5]}] (0) --(3);
\draw[black!70, dashed, rounded corners=2pt] (-1.55,0.25) rectangle (-0.3,1.25);
\draw[black!70, dashed, rounded corners=2pt] (1.55,0.25) rectangle (0.3,1.25);
\end{tikzpicture}
\caption{Illustration of dyadic outcomes and unit-level outcomes. The upstream and downstream neighbour sets for unit $i=1$ are $\mathcal{U}_1 = \{2,3\}$ and $\mathcal{D}_1 = \{3,4\}$, respectively. }\label{fig:example}
\end{figure}

Although the outcome of interest is dyadic, we consider a standard unit-level randomization setting, which is pervasive in online controlled experiments. Let $\bm{Z}=(Z_{1},Z_{2},\dots,Z_{n})\in \{0,1\}^n$ denote a vector of binary treatments assigned to the units, where $Z_i=1$ if unit $i$ receives the treatment and $Z_i=0$ if unit $i$ receives the control. In a randomized experiment, the treatment vector $\bm{Z}=\bm z$ is assigned according to a predetermined distribution $\pr(\bm z)$. Although dyad-level randomization is conceptually possible, assigning treatment separately to each pairwise interaction is often operationally undesirable. For example, a single user could simultaneously experience both treatment and control interfaces or features across different peers, which may alter user experience and compromise the interpretation of the experimental contrast. We therefore focus on the standard setting in which treatment is assigned at the unit level.

Following the potential outcome framework, we let $Y_{ij}(\bm z)$ denote the potential dyadic outcome that would be observed if the assignment vector were set to $\bm Z=\bm z$. In the presence of interference, $Y_{ij}(\bm z)$ may depend on the entire assignment vector. We adopt a design-based finite-population framework \citep{neyman1923applications,rubin1974estimating,imbens2015causal}: all potential dyadic outcomes are fixed, and the only source of randomness is the randomized assignment.
{The design-based formulation identifies the source of uncertainty and avoids specifying model components needed for estimation and inference under interference. Rather than specifying an outcome distribution, an exposure model, or a network-generating process, randomization-based inference conditions on the realized potential-outcome table and uses the known assignment mechanism as the sole source of randomness. For this reason, a finite-population or design-based perspective is widely used in the modern interference literature \citep{hudgens2008interference,aronow2017estimating,savje2021average,savje2024causal}. Our work follows this tradition and extends it from unit-level to dyadic outcomes.}

For each assignment vector $\bm z$, define the potential neighbour sets $\mathcal{U}_i(\bm z)=\{j:Y_{ji}(\bm z)\neq\dagger\}$, $\mathcal{D}_i(\bm z)=\{j:Y_{ij}(\bm z)\neq\dagger\}$, and $\mathcal{N}_i(\bm z)=\mathcal{U}_i(\bm z)\cup\mathcal{D}_i(\bm z)$. The corresponding unit-level potential outcomes $U_i(\bm z)$, $D_i(\bm z)$, and $Y_i(\bm z)$ are defined by applying \eqref{eq:outcomes} to these potential neighbour sets and potential dyadic outcomes. {We allow the neighbour sets to vary with treatment status, which differs from the majority of existing literature that assumes a fixed interference network before and after the experiments. Recent work also allows treatment-dependent interaction structures in unit-level outcome models, for example through mis-specified exposure mapping \citep{savje2024causal} or  instrumental variables based on neighbourhood treatment \citep{gao2024endogenous, gao2024dissertation, shankar2025experimentation}. Our setting differs in focusing on dyadic potential outcomes and randomization-based inference, with the assignment mechanism as the only source of stochastic variation.}

Consider the two counterfactual regimes in which all units receive treatment or all units receive control. The global average treatment effect for unit-level outcomes is defined as
\begin{equation}\label{eq:tte}
    {\tau} = \frac{1}{n}\sum_{i=1}^n \{U_{i}(\bm{1}) - U_{i}(\bm{0})\} = \frac{1}{n}\sum_{i=1}^n \{D_{i}(\bm{1}) - D_{i}(\bm{0})\},
\end{equation}
where $\bm{1}, \bm{0}$ are the length-$n$ vectors of ones and zeros, respectively. 
This estimand compares the average unit-level potential outcome under the hypothetical regime in which all users adopt the treatment with that under the regime in which all users remain under control.
It can be  equivalently defined with dyadic outcomes as a contrast of the corresponding dyadic interactions under the two global treatment regimes, 
\[
{\tau} = \frac{1}{n}\sum_{i=1}^n \Big( \sum_{j\in \mathcal{U}_{i}(\bm{1})} Y_{ji}(\bm{1}) - \sum_{j\in \mathcal{U}_{i}(\bm{0})} Y_{ji}(\bm{0})\Big)
= \frac{1}{n}\sum_{i=1}^n \Big( \sum_{j\in \mathcal{D}_i(\bm{1})} Y_{ij}(\bm{1}) - \sum_{j\in \mathcal{D}_i(\bm{0})} Y_{ij}(\bm{0})\Big).
\]
Beyond the global average treatment effect $\tau$, other causal estimands may also be of interest, such as direct, indirect, and spillover effects \citep{vanderweele2011effect, savje2021average, hu2022average}. Under the present dyadic-outcome framework, these estimands can be represented in an analogous dyad-level form. In this paper, we focus primarily on inference for the global average treatment effect in \eqref{eq:tte}.

As established by \citet{basse2018limitations}, the global treatment effects are generally not identifiable without restrictions on the interference structure. The same challenge arises for dyadic outcomes if $Y_{ij}(\bm z)$ is allowed to depend arbitrarily on the entire assignment vector. We therefore impose the following dyadic interference restriction.
\begin{assumption}[Dyadic interference]\label{asmp:pairitf}
$Y_{ij}(\bm{z}) = Y_{ij}(z_i, z_j)$ for $i\neq j$. In the main analysis, we set $Y_{ii}(\bm z)=\dagger$ for all $i$ and $\bm{z}$.
\end{assumption}

Assumption~\ref{asmp:pairitf} states that each dyadic outcome depends on treatment only through the two endpoints of the dyad.
{This condition is applicable for outcomes generated by a direct pairwise interaction, especially when the outcome is measured over a short horizon. Examples include VoIP call duration or connection quality, pairwise transaction outcomes, and direct-message responses.

When dyadic interactions are the primary source of treatment effect but higher-order propagation may still occur, Assumption~\ref{asmp:pairitf} serves as a useful working approximation, for example in short-horizon within-session engagements or dwell time on a shared page. It may be less appropriate for outcomes driven by long-horizon and higher-order spillovers, or engagement feedback loops, where treatment can propagate beyond the focal dyad through subsequent interactions. Thus, the assumption is intended to formalize a tractable dyadic dependence structure for outcomes whose main treatment dependence is pairwise, rather than to capture all possible network dynamics. We assess robustness to potential violations through the sensitivity analyses in Section~\ref{sec:dis} and Section~C.2 of the Supplementary Material.}

{Under the design-based perspective, the dyadic potential outcomes are fixed finite-population quantities, with the treatment assignment serving as the sole source of randomness. We therefore do not impose independence, exchangeability, a superpopulation model, or a parametric outcome model on $Y_{ij}(z_i,z_j)$. Dependence arising from shared units and heterogeneous degrees is instead captured by the degree-related quantities that govern the asymptotic analysis in Section~\ref{sec:asymp}. This contrasts with model-based approaches to dyadic data, which typically specify a stochastic structure for dyadic outcomes and use dyadic-robust variance corrections to account for dependence among dyads sharing a unit \citep{fafchamps2007risk,tabord2019inference,graham2020dyadic}.}

Assumption \ref{asmp:pairitf} only restricts the interference structure on the dyadic outcomes;
however, under this assumption, the interference structure on the unit-level outcomes can still be complex.
{First, the set of units whose treatments can affect a given unit's unit-level outcome, referred to as the unit-level interference set, can be much larger than the pair of units involved in a dyadic outcome. Formally, under Assumption~\ref{asmp:pairitf}, the unit-level potential outcomes $U_i(\bm{z}), D_i(\bm{z})$ and $Y_i(\bm{z})$ depend on treatments of unit $i$ together with all units in $\mathcal{U}_i(\bm{z})$, $\mathcal{D}_i(\bm{z})$, and $\mathcal{N}_i(\bm{z})$, respectively. The cardinality of this set can grow with the sample size $n$, whereas at the dyad level only two treatment indicators $(z_i,z_j)$ are relevant. 
Hence, dyadic interference implies neighbourhood interference for unit-level outcomes,
where a unit's potential outcome depends on the treatments of its neighbours \citep[see][]{eckles2016design,aronow2017estimating,yu2022estimating}. 
However, unlike the previous neighbourhood interference defined on a static network, the set of neighbours also depends on the treatment status in our setting. 
Second, unit-level interference effects can be heterogeneous across neighbours. } Under Assumption~\ref{asmp:pairitf} and the zero convention ($\dagger = 0$) for absent dyads,
\[
\begin{aligned}
    U_{i}(\bm{z}) &= \sum_{j\neq i} Y_{ji}(z_j,z_i)
    = U_i(\bm 0)+\alpha_i z_i+\sum_{j\neq i}(\eta_{ji}z_j+\lambda_{ji}z_iz_j),\\
    D_{i}(\bm{z}) &= \sum_{j\neq i} Y_{ij}(z_i,z_j)
    = D_i(\bm 0)+\beta_i z_i+\sum_{j\neq i}(\gamma_{ij}z_j+\lambda_{ij}z_iz_j),
\end{aligned}
\]
where $\alpha_i=\sum_{j\neq i}\{Y_{ji}(0,1)-Y_{ji}(0,0)\}$ and $\beta_i=\sum_{j\neq i}\{Y_{ij}(1,0)-Y_{ij}(0,0)\}$ are direct effects on unit $i$ through upstream and downstream aggregations, respectively. The coefficients $\eta_{ji}=Y_{ji}(1,0)-Y_{ji}(0,0)$ and $\gamma_{ij}=Y_{ij}(0,1)-Y_{ij}(0,0)$ capture indirect dyad-level effects, and $\lambda_{ij}=Y_{ij}(1,1)-Y_{ij}(1,0)-Y_{ij}(0,1)+Y_{ij}(0,0)$ captures treatment interaction within dyad $(i,j)$. The dyadic treatment effect is $\tau_{ij}=Y_{ij}(1,1)-Y_{ij}(0,0)$, and the global average treatment effect can be written as $\tau=n^{-1}\sum_{i=1}^n\sum_{j\neq i}\tau_{ij}$. Heterogeneity is allowed because these coefficients may vary freely across units and dyads.
This representation also clarifies the connection with heterogeneous additive network-effect models. For example, \citet{yu2022estimating} considered an additive specification with $\lambda_{ij}=0$ and showed that additional information from historical data or pilot studies is needed for unbiased estimation of the global average treatment effect. 
However,  Assumption \ref{asmp:pairitf} not only admits neighbourhood interference with a possibly varying network structure but also allows for heterogeneity of unit-level effects without any specific exposure mapping.

 We also impose a common positivity assumption on the treatment assignment in randomized experiments.
Let $p_{i} = \pr(Z_{i}=1)$ and $q_i = \pr(Z_i=0)$, and define $P_1 = \min_{1\leq i\leq n}p_i$ and $P_0 = \min_{1\leq i\leq n} q_i$ as the minimum of treatment and control probabilities, respectively.
\begin{assumption}[Positivity]\label{asmp:positivity}
$P_1>0$ and $P_0 > 0$ for all $n$.
\end{assumption}

Assumption \ref{asmp:positivity} requires that the unit-level treatment and control probabilities are positive. Note that $P_1, P_0$ can depend on $n$ in some randomization designs. For example, under complete randomization,  $P_1 = n_1/n$ with $n_1$ being the number of treated units that is determined in advance. When needed below, we explicitly impose that they are uniformly bounded away from zero and one.

To estimate $\tau$, one approach is to leverage the unit-level outcomes by  constructing a class of Horvitz-Thompson type  estimators,
\begin{equation}\label{eq:unit_weight}
    \tilde{\tau}(\bm{Y}) = \frac{1}{n}\sum_{i=1}^{n} \left(\frac{Z_i}{a_i} - \frac{1-Z_i}{b_i}\right)Y_i,
\end{equation}
where $a_i, b_i$ denote some pre-specified weights for units. Alternatively, one could also use either upstream or downstream unit-level outcomes and construct estimators $\tilde{\tau}(\bm{U})$ and $\tilde{\tau}(\bm{D})$. 
Under Assumption \ref{asmp:pairitf}, we have the following result.
\begin{theorem}\label{thm:nodestimator}
    {Under Assumptions~\ref{asmp:pairitf}--\ref{asmp:positivity}, (i) For any pre-specified weights $a_i,b_i\in(0,1)$, the estimators $\tilde{\tau}(\bm U)$ and $\tilde{\tau}(\bm D)$ are not uniformly unbiased for $\tau$ over the class of potential outcomes satisfying Assumption~\ref{asmp:pairitf}; (ii) the estimator $\tilde\tau(\bm Y)$ is uniformly unbiased for $\tau$ over the same class if and only if $p_i=a_i=b_i=1/2$, for $i=1,\ldots,n.$ }
\end{theorem}

{Theorem~\ref{thm:nodestimator} shows the weighted estimators based on upstream and downstream aggregated outcomes are not uniformly unbiased, and the total unit-level estimator is uniformly unbiased only under the balanced independent Bernoulli design with matching weights. Outside this special case, one can construct non-degenerate sequences of potential outcomes for which the unit-level estimator has non-vanishing bias. The failure is not caused by aggregation, but by the fact that aggregation mixes dyads with different endpoint treatment configurations, thereby changing the exposure structure relative to the global average treatment effect. Similarly, under Assumptions~\ref{asmp:pairitf}--\ref{asmp:positivity}, one can also construct sequences of potential outcomes for which $\tilde\tau(\bm U)$ and $\tilde\tau(\bm D)$ are inconsistent for $\tau$.}
Result (i) is an extension of the results by \cite{yu2022estimating} under the heterogeneous additive network effect. 
The condition in Result~(ii) is therefore a knife-edge case: it requires balanced independent Bernoulli assignment together with matching unit weights. 
In the next section, we resort to using dyadic data for the estimation of the global average treatment effect $\tau$.

\section{Estimation of the global average treatment effect with dyadic data}\label{sec:estimator}
\subsection{Estimators}
We consider using dyadic outcomes for the estimation of the global average treatment effect $\tau$.
Let $Z_{ij} = Z_{i}Z_{j}$ and $\overline{Z}_{ij} = (1-Z_i)(1-Z_j)$ be the indicators that units $(i,j)$ both receiving the treatment or the control, respectively. 
Suppose $p_{ij} = E(Z_{ij})$ and $q_{ij} = E(\overline{Z}_{ij})$ are known under the randomization design. {Throughout the theoretical statements below, the dyad-level inclusion probabilities $p_{ij}$ and $q_{ij}$ appearing in inverse-probability weights are assumed to be strictly positive; under complete randomization this requires at least two treated/control units. Under Assumption~\ref{asmp:pairitf}, an observed dyad contributes information about the all-treated potential outcome only when both endpoints are treated, and about the all-control potential outcome only when both endpoints are under control. }
Motivated by this, we first propose a Horvitz-Thompson type estimator,
\begin{equation}
\begin{aligned}\label{eq:tauhat}
\hat{\tau}_{\rm HT}= \frac{1}{n}\sum_{i=1}^n \sum_{j\in \mathcal{D}_i}\left(\frac{Z_{ij}}{p_{ij}} - \frac{\overline{Z}_{ij}}{q_{ij}}\right)Y_{ij}.
\end{aligned}
\end{equation}
This estimator is a contrast of two parts, the first part is a weighted average of the dyadic outcomes for unit pairs that   both receive the treatment,
and the second part is a weighted average of the dyadic outcomes for unit pairs that both receive the control.
The estimator accommodates possibly unequal assignment probabilities, which allows for general designs $\pr(\bm{z})$. 
In addition, a H\'{a}jek type estimator \citep{hajek1971comment} is proposed
\begin{equation}\label{eq:tauha}
\begin{aligned}
    \hat{\tau}_{\rm H\acute{A}} =&  \frac{1}{\hat{n}_1}\sum_{i=1}^n \sum_{j\in \mathcal{D}_i}  \frac{Z_{ij}Y_{ij}}{p_{ij}} -  \frac{1}{\hat{n}_0}\sum_{i=1}^n \sum_{j\in \mathcal{D}_i} \frac{\overline{Z}_{ij}Y_{ij}}{q_{ij}}, 
\end{aligned}
\end{equation}
where $\hat{n}_1 = \sum_{i=1}^n  p_{i}^{-1}Z_{i}$ and $\hat{n}_0 = \sum_{i=1}^n  q_i^{-1}(1-Z_{i})$. 
The H\'{a}jek estimator normalizes the inverse-probability weighted sums and can be more stable in finite samples, although it is not guaranteed to be uniformly more efficient than the Horvitz--Thompson estimator. This property is well known for estimators based on unit-level outcomes in the survey-sampling literature \citep{sarndal2003model}. We further illustrate this with numerical examples in Section \ref{sec:simu1}.

{
For both estimators, the summation is taken over the dyads observed under the realized assignment. This differs from conditioning on a post-treatment selected subset, such as earnings conditional on choosing to work. Rather, the estimator targets a contrast between two complete counterfactual regimes, allowing both the magnitudes of dyadic outcomes and the induced interaction patterns to differ across assignments. Equivalently, one may view the summation as being taken over all ordered dyads $(i,j)$, with non-interacting dyads contributing zero. The observed neighbour sets provide a compact representation of this sparse dyadic structure. This formulation covers settings with a known social network $\mathcal G$, where potential dyads are restricted to a pre-specified edge set $\mathcal E$, as well as settings without such a network, where treatment may affect whether users forward, view, or otherwise interact with one another. If the scientific question concerns effects on a fixed set of dyads, the same estimators can be applied after restricting to that pre-specified edge set $\mathcal E$. Our framework is also related to recent work on treatment-induced network changes. For example, \citet{gao2024endogenous} studies unit-level outcomes and treats post-treatment network changes as a mediation channel. In contrast, we study dyadic outcomes $Y_{ij}$ directly: treatment-dependent interaction sets are part of the potential dyadic outcome structure rather than a mediator to be decomposed. Our inference is design-based and relies only on known assignment probabilities, so the two approaches are complementary.}

\subsection{Asymptotic properties}\label{sec:asymp}

In this section, we focus on the asymptotic properties of the proposed estimators \eqref{eq:tauhat} and \eqref{eq:tauha}. 
Under Bernoulli randomization, the assignment probability is ${\rm pr}(\bm{z})=\prod_{i=1}^n p_{i}^{z_i}q_{i}^{1-z_i}$, where each $Z_i$ independently follows ${\rm Bernoulli}(p_i)$ for prescribed probabilities $p_i$ for $1\leq i\leq n$. This design is common in online controlled experiments. 
Under complete randomization, the number of treated units is fixed at $n_1$, and all assignments with exactly $n_1$ treated units are equally likely: 
${\rm pr}(\bm{z})= \binom{n}{n_1}^{-1}  \text { if } \sum_{i=1}^n z_i=n_1$ and 
${\rm pr}(\bm{z})= 0$ otherwise. 
Even though treatment is assigned at the unit level, the estimators aggregate dyadic outcomes, creating dependence among dyads that share a unit. Consequently, the convergence rate and asymptotic distribution depend on both the randomization design and the complexity of dyadic interaction structure. The Supplementary Material further extends the analysis to cluster randomization, which is useful when treatment is assigned at the level of groups rather than individual units.
To quantify this complexity, we introduce degree-related quantities defined from the potential neighbour sets. For a set $\mathcal S$, let $|\mathcal S|$ denote its cardinality. 

\begin{definition}\label{def:avg}
For $\mathsf z\in\{0,1\}$, let $\bm{\mathsf z}=(\mathsf z,\ldots,\mathsf z)$ denote the assignment vector of length $n$ under which all units receive treatment status $\mathsf z$. Define
\begin{itemize}
\item the average counterfactual degree: $d_{1}(\mathsf{z}) = n^{-1}\sum_{i=1}^n |\mathcal{N}_{i}(\bm{\mathsf{z}})|$;
\item {the mean squared counterfactual degree:} $
d_{2}(\mathsf{z}) = n^{-1}\sum_{i=1}^n |\mathcal{N}_{i}(\bm{\mathsf{z}})|^2;
$
\item the maximum counterfactual degree: $
d_{\infty}(\mathsf{z}) = \max_{1\leq i\leq n} |\mathcal{N}_{i}(\bm{\mathsf{z}})|.
$
\end{itemize}
\end{definition}

These quantities summarize the complexity of the two counterfactual interaction networks induced by the potential dyadic outcomes under the all-treated and all-control assignments. 
Specifically, $d_1(\mathsf z)$ measures the average degree of the counterfactual network under $\bm{\mathsf z}$, while $d_2(\mathsf z)$ is the mean squared counterfactual degree and is related to the number of pairs of dyadic outcomes that share a common unit. The quantity $d_\infty(\mathsf z)$ measures the largest counterfactual degree and controls the influence of highly connected units. By Jensen's inequality and the bound $|\mathcal N_i(\bm{\mathsf z})|\leq d_\infty(\mathsf z)$, we have $d_1(\mathsf z)^2\leq d_2(\mathsf z)\leq d_1(\mathsf z)d_\infty(\mathsf z).$

\begin{assumption}\label{asmp:bound}
    The potential dyadic outcomes are uniformly bounded: for all $n\geq 2$, $\max_{1\leq i,j\leq n}\max_{\bm{z}\in\{0,1\}^n} |Y_{ij}(\bm{z})| \leq K$ for some constant $K>0$. 
\end{assumption}

Assumption~\ref{asmp:bound} is a standard regularity condition in design-based causal inference. It can be relaxed under suitable moment conditions, but it is sufficient for the asymptotic results developed below. Importantly, boundedness of dyadic outcomes is weaker than boundedness of unit-level aggregate outcomes, because the number of neighbours in $\mathcal N_i(\bm z)$ may grow with $n$. The assumption is also natural in many online experiments: click indicators are binary dyadic outcomes, and VoIP call durations are bounded continuous outcomes over a fixed observation window.

\begin{theorem}\label{thm:rate}
Under Assumptions \ref{asmp:pairitf}-\ref{asmp:bound}, 
with Bernoulli randomization or complete randomization, (i). the estimator $\hat{\tau}_{\rm HT}$ is unbiased; (ii).  
$
    \hat{\tau}_{\rm HT}-\tau =  {O}_{p}\left(r_n\right)$ and $\hat{\tau}_{\rm H\acute{A}}-\tau= {O}_{p}\left(r_n\right) 
$
where \[
    r_n = \max_{\mathsf{z}\in\{0,1\}}\left\{P_{\mathsf{z}}^{-1}d_{1}(\mathsf{z})^{1/2} , P_{\mathsf{z}}^{-1/2}d_{2}(\mathsf{z})^{1/2} \right\} n^{-1/2}.
    \]
\end{theorem}
Hereafter, ${O}_{p}(1)$ denotes boundedness in probability, and $o_p(1)$ denotes convergence to zero in probability.
Theorem \ref{thm:rate} establishes the convergence rate of estimators based on dyadic outcomes. The rate depends on both the treatment assignment probabilities and the complexity of the counterfactual interaction networks. 
The factor $P_{\mathsf z}$ is retained in the bound to allow treatment or control probabilities to decrease with $n$. This accommodates practical experimental designs in which one arm may receive only a small fraction of traffic.
For example, a costly or risky treatment may be tested on a small treated group, corresponding to a small $P_1$, while a post-launch holdout experiment may use a small control group, corresponding to a small $P_0$. The theorem shows that, despite the additional dependence induced by complete randomization, the same order of convergence is obtained under Bernoulli and complete randomization.

The degree-related terms capture the effect of dyadic dependence. The term $d_1(\mathsf z)^{1/2}P_{\mathsf z}^{-1}$ reflects the contribution of the average counterfactual degree, whereas $d_2(\mathsf z)^{1/2}P_{\mathsf z}^{-1/2}$ reflects the contribution of dyads sharing common units. Larger values of $d_1(\mathsf z)$ or $d_2(\mathsf z)$ indicate stronger interaction structure and lead to slower convergence.

To illustrate the rate, suppose that all units have the same  counterfactual degree, i.e., $|\mathcal N_i(\bm{\mathsf z})|=d$, and that the treatment probabilities are homogeneous. Then $d_1(\mathsf z)=d$ and $d_2(\mathsf z)^{1/2}=d$, so the rate becomes
$n^{-1/2}\max\{d^{1/2}(P_1P_0)^{-1}, d(P_1P_0)^{-1/2}\}.$ If the treatment and control probabilities are bounded away from zero, this reduces to $O(dn^{-1/2})$. Hence, the estimators are root-$n$ consistent when the degree $d$ is bounded, while the rate slows proportionally to $d$ when the number of neighbours grows with $n$. When one assignment probability is small, the convergence rate is further inflated by the corresponding inverse-probability factors.

{If the potential dyadic outcomes are instead viewed as random draws from a superpopulation, the design-based results can be interpreted conditionally on the realized potential outcomes. In particular, provided that the treatment assignment is independent of the potential-outcome generating process, the Horvitz--Thompson estimator remains unbiased for the realized finite-population estimand and hence is also unbiased for the corresponding superpopulation estimand after averaging over the outcome-generating process. The main difference concerns variance interpretation. By the law of total variance, the unconditional variance equals the expected design variance plus the additional variation of the realized finite-population estimand across draws of the potential outcomes. The variance estimators developed below target the conditional design variance; inference for a fully unconditional superpopulation estimand would require modelling or estimating this additional component, which is outside the scope of the present design-based analysis.}

In the remainder of the paper, we assume that the treatment and control probabilities are bounded away from zero, i.e., $\min(P_1,P_0)>c$ for some constant $c>0$. This condition 	holds when treatment and control fractions are non-vanishing and simplifies the rate in Theorem~\ref{thm:rate}.
\begin{theorem}\label{thm:p}
    For Bernoulli randomization or complete randomization with $\min(P_1,P_0)>c$ for some $ c>0$, under Assumptions \ref{asmp:pairitf} and \ref{asmp:bound}, 
    we have 
    $
    \hat{\tau}_{\rm HT}-\tau = {O}_{p}\left(\max_{\mathsf{z}\in\{0,1\}} d_{2}(\mathsf{z})^{1/2} n^{-1/2} \right)
    $ and $\hat{\tau}_{\rm H\acute{A}}-\tau = {O}_{p}\left(\max_{\mathsf{z}\in\{0,1\}} d_{2}(\mathsf{z})^{1/2} n^{-1/2} \right).$
\end{theorem}
Theorem \ref{thm:p} shows that, when the assignment probabilities are bounded away from zero, the convergence rate is governed by the mean squared counterfactual degree. 
This quantity captures not only the overall density of the interaction network but also the imbalance of the degree distribution. Thus, highly connected units can slow convergence even when the average degree is small. 
A star network illustrates this point. Suppose one central unit is connected to all other units, while the remaining units are mutually disconnected. Then $d_1(\mathsf z)<2$, but $d_2(\mathsf z)=n-1$, so the rate bound in Theorem~\ref{thm:p} is of constant order. 
{Under Bernoulli randomization, if all observed dyads share the central unit, the leading component of $\hat{\tau}_{\rm HT}$ depends on the treatment assignment of that central unit, i.e.,
\[
\hat{\tau}_{\rm HT} = \frac{1}{n}\sum_{j\neq1}\left(\frac{Z_{j1}Y_{j1}}{p_{j1}}-\frac{\overline{Z}_{j1}Y_{j1}}{q_{j1}}\right).
\]
Consequently, $\hat{\tau}_{\rm HT}$ converges to a non-degenerate  function of $Z_1$ rather than converging in probability to $\tau$. The failure arises because the dyadic information is concentrated around a single unit $i=1$.}

The rate in Theorem~\ref{thm:p} is analogous to rates for unit-level estimators under interference,  where dependence is also governed by the structure of exposure mappings or neighbourhoods \citep[see Propositions~2--3 of][]{savje2021average}.
A key distinction is that the proposed $\hat{\tau}_{\rm HT}$ and $\hat{\tau}_{\rm H\acute{A}}$ are asymptotically unbiased for $\tau$, whereas the corresponding unit-level estimator $\tilde{\tau}(\bm{Y})$ is generally biased under dyadic interference.

We next move from convergence rates to asymptotic distributions. The following condition controls the extent to which dependence can concentrate around highly connected units.
\begin{assumption}\label{asmp:netrate} 
{Suppose the estimators have positive variance $\sigma_{\rm HT}^2=\operatorname{var}(\hat{\tau}_{\rm HT}) > 0$ and $\sigma_{\rm H\acute{A}}^2=\operatorname{var}(\hat{\tau}_{\rm H\acute{A}}) > 0$, and 
$\max_{\mathsf{z}\in\{0,1\}}d_{\infty}(\mathsf{z})n^{-2/3} \max\{\sigma_{\rm HT}^{-1},\sigma_{\rm H\acute{A}}^{-1}\} = o(1)$.}
\end{assumption}

\begin{theorem}\label{thm:clt}
    For Bernoulli randomization with $\min(P_1,P_0)>c$ for some $c>0$, under Assumptions \ref{asmp:pairitf}-\ref{asmp:netrate}, we have
    $
   \sigma_{\rm HT}^{-1}(\hat\tau_{\rm HT} - \tau) \stackrel{d}{\rightarrow} N(0,1)
    $
    and 
    $
    \sigma_{\rm H\acute{A}}^{-1}(\hat\tau_{\rm H\acute{A}} - \tau) \stackrel{d}{\rightarrow} N(0,1). 
    $
\end{theorem}

{Assumption~\ref{asmp:netrate} is a design-based regularity condition that limits the influence of highly connected units relative to the variance scale of the estimator. It does not impose independence, exchangeability, or a superpopulation model on the dyadic potential outcomes. Instead, it restricts the realized potential-outcome networks through the maximum counterfactual degree $d_\infty(\mathsf z)$ and the randomization variance of $\hat\tau_{\rm HT}$. Its role is to rule out configurations in which a small number of units dominate the assignment-induced dependence. Similar maximum-degree conditions appear in dyadic regression and in central limit theorems for graph-dependent random variables \citep[e.g.,][]{janson1988normal, tabord2019inference}.}

{Theorem \ref{thm:clt} highlights that asymptotic normality is compatible with growing degrees, provided that the interaction network is sufficiently balanced relative to the estimator's variance. 
In dense but balanced networks, each dyad shares endpoints with only a vanishing fraction of all ordered dyads, so no single unit dominates the randomization-induced dependence. 
By contrast, normality may fail under degree concentration, as in the star-network example above, where the estimator is driven by the treatment assignment of the central unit and converges to a non-degenerate non-normal limiting distribution. } The proof for $\hat\tau_{\rm HT}$ under Bernoulli randomization uses Stein's method for locally dependent random variables \citep{ross2011fundamentals}.

\section{Variance and confidence interval}\label{sec:confidence}

To quantify uncertainty and construct confidence intervals for the global average treatment effect, we need estimators of the asymptotic variance. 
In design-based causal inference, however, exact variance estimation is often difficult because the variance generally depends on unobserved potential outcomes \citep{imbens2015causal}. This difficulty is further amplified under interference, where the estimator involves dependent potential outcomes across units or dyads \citep[Section~6]{savje2021average}.
We therefore seek variance estimators that are conservative, or conservative up to asymptotically negligible error, under some degree conditions. 
Denote $S_{ij} = Y_{ij} +Y_{ji}$ and $S_{ij}(z_i,z_j) = Y_{ij}(z_i,z_j) +Y_{ji}(z_j,z_i)$. Under Bernoulli randomization, the variance of the estimator $\hat{\tau}_{\rm HT}$ can be written as
\begin{equation}\label{eq:vartau}
\begin{aligned}  
\sigma^2_{\rm HT} = &
\frac{1}{n^2}\sum_{i=1}^n \sum_{j\neq i} \bigg\{\frac{1}{2}\left( \frac{S_{ij}(1,1)^2}{p_{ij}} + \frac{S_{ij}(0,0)^2}{q_{ij}} - (\tau_{ij} + \tau_{ji})^2 \right)\\
+ &\sum_{\substack{\substack{k\neq i\\k\neq j}}} \left(\frac{S_{ij}(1,1)S_{ik}(1,1)}{p_i} + \frac{S_{ij}(0,0)S_{ik}(0,0)}{q_i}\right) - (\tau_{ij}+\tau_{ji})(\tau_{ki} + \tau_{ik}) \bigg\}.
\end{aligned}
\end{equation} 
The first line contains dyad-level variance terms, whereas the second line contains covariance terms between dyads that share unit $i$. These triadic terms arise because outcomes such as $Y_{ij}$ and $Y_{ik}$ depend on the common treatment assignment $Z_i$.
We start with a plug-in variance estimator, 
\begin{equation}\label{eq:varest}
\begin{aligned}  
\hat{\sigma}^2 =& \frac{1}{n^2}\sum_{i=1}^n \sum_{j\neq i} \bigg\{\frac{1}{2} \left( \frac{Z_{ij}}{p_{ij}^2} + \frac{\overline{Z}_{ij}}{q_{ij}^2} \right) S_{ij}^2 +  \sum_{\substack{k\neq i\\k\neq j}} \left( \frac{Z_{ijk}}{p_{ij}p_{ik}} + \frac{\overline{Z}_{ijk}}{q_{ij}q_{ik}} \right) S_{ij}S_{ik}\bigg\},
\end{aligned}
\end{equation}
where $Z_{ijk}=Z_iZ_jZ_k$ and $\overline{Z}_{ijk}=(1-Z_i)(1-Z_j)(1-Z_k)$ indicate that triads $(i,j,k)$ are all assigned to treatment or control, respectively. 
The variance estimator $\hat{\sigma}^2$ and the true variance $\sigma^2_{\rm HT}$ both contain two types of terms: variance and covariance terms within each reciprocal dyad pair $(Y_{ij},Y_{ji})$, and covariance terms between $Y_{ij}$ and other dyadic outcomes that share a common unit index. The estimator $\hat{\sigma}^2$ replaces the counterfactual outcomes with its weighted observable counterparts, e.g., replacing $p^{-1}_{ij}S_{ij}(1,1)^2$ with $p_{ij}^{-2}Z_{ij}S_{ij}^2$. {Although \eqref{eq:varest} follows the structure of the true variance, it is not generally unbiased: interference creates covariance terms involving multiple unobserved potential outcomes, and their plug-in replacements can have non-negligible asymptotic bias. 
The following condition controls this bias by ruling out extreme concentration of degree mass. }
\begin{assumption}\label{asmp:var}
    $\max_{\mathsf{z}\in\{0,1\}}\{d_{\infty}(\mathsf{z})d_{2}(\mathsf{z})^{-1/2} \}n^{-1/2} = o(1)$.
\end{assumption}

Assumption \ref{asmp:var} is a balance condition on the counterfactual degree distributions under the all-treated and all-control regimes.
{It requires the maximum counterfactual degree to be small relative to the aggregate degree variation measured by $d_2(\mathsf z)^{1/2}n^{1/2}$. The condition holds  in balanced networks. For example, if all units have the same degree $|\mathcal N_i(\bm{\mathsf z})|=d$, then $d_\infty(\mathsf z)=d$ and $d_2(\mathsf z)=d^2$, so the left-hand side is $O(n^{-1/2})$. By contrast, the condition fails in a star network: $d_\infty(\mathsf z)=n-1$ and $d_2(\mathsf z)=n-1$, making the left-hand side of constant order. Thus, Assumption~\ref{asmp:var} excludes configurations in which a small number of highly connected units dominate the variance estimation problem. }  Recall that $\tau_{ij} = Y_{ij}(1,1) - Y_{ij}(0,0) $ represents the dyadic treatment effect.

\begin{theorem}\label{thm:varbias}
For Bernoulli randomization with $\min(P_1,P_0)>c$ for some $c>0$, under Assumptions \ref{asmp:pairitf}-\ref{asmp:bound} and \ref{asmp:var}, we have 
\[
\max_{\mathsf{z}\in\{0,1\}}nd_{2}(\mathsf{z})^{-1} \{\hat{\sigma}^2 -\sigma^2_{\rm HT} - {\rm bias}(\hat{\sigma}^2) \} =o_p(1),
\]
where ${\rm bias}(\hat{\sigma}^2)= n^{-2}\sum_{i=1}^n \sum_{j\neq i} \{(\tau_{ij} + \tau_{ji})^2 / 2+ \sum_{\substack{k\neq i\\k\neq j}}  (\tau_{ij} + \tau_{ji})(\tau_{ik} + \tau_{ki})\}.$
\end{theorem}

Theorem~\ref{thm:varbias} shows that the plug-in estimator $\hat{\sigma}^2$ differs from the true randomization variance by an explicit asymptotic bias term. The bias consists of products of dyad-level treatment effects. The first component involves squared symmetrized dyadic effects, while the second component involves products of effects for dyads sharing a common unit.
If all dyadic treatment effects are non-negative, then ${\rm bias}(\hat{\sigma}^2)\geq0$ and $\hat{\sigma}^2$ is asymptotically conservative.
In general, however, dyadic effects may have mixed signs, so the bias need not be non-negative and $\hat{\sigma}^2$ is not guaranteed to be conservative. 

To address the possible non-conservativeness of $\hat{\sigma}^2$, we construct an adjusted variance estimator by augmenting  $\hat{\sigma}^2$ with an additional parameter $\Delta$:
\begin{equation}\label{eq:varest2}
\begin{aligned}  \hat{\sigma}^2(\Delta) =& \frac{1}{n^2}\sum_{i=1}^n \sum_{j\neq i} \bigg\{ \frac{1}{2}\bigg(\frac{Z_{ij}}{p_{ij}^2} + \frac{\overline{Z}_{ij}}{q_{ij}^2}\bigg)S_{ij}^2 + \Delta\bigg(\frac{Z_{ij}}{p_{ij}} + \frac{\overline{Z}_{ij}}{q_{ij}}\bigg)S_{ij}^2 + \sum_{\substack{k\neq i\\k\neq j}} \bigg( \frac{q_iZ_{ijk}}{p_{ij}p_{ik}} + \frac{p_i\overline{Z}_{ijk}}{q_{ij}q_{ik}} \bigg) S_{ij}S_{ik} \bigg\}.
\end{aligned}
\end{equation}
{The parameter $\Delta$ controls the magnitude of degree-based adjustment for potentially non-conservative covariance terms.}
This variance estimator \eqref{eq:varest2} is obtained by bounding the product of counterfactual outcomes in \eqref{eq:varest} with Young's inequality. 
Specifically, ${\rm bias}(\hat{\sigma}^2)$ involves the summation of product terms, such as $\tau_{ij}\tau_{ik} = Y_{ij}(1,1)Y_{ik}(1,1) - Y_{ij}(1,1)Y_{ik}(0,0)- Y_{ij}(0,0)Y_{ik}(1,1) + Y_{ij}(0,0)Y_{ik}(0,0)$. 
The middle two terms $Y_{ij}(1,1)Y_{ik}(0,0)$ and $Y_{ij}(0,0)Y_{ik}(1,1)$ involve products of potential outcomes under different treatment regimes and cannot be observed jointly. Young's inequality gives, for example, $Y_{ij}(1,1)Y_{ik}(0,0) \leq \{Y_{ij}(1,1)^2 + Y_{ik}(0,0)^2\}/2, $ and analogous bounds apply to the other cross-world products. Each squared counterfactual outcome is then replaced by its inverse-probability-weighted observed counterpart, such as $Y_{ij}(1,1)^2
\leadsto p_{ij}^{-1}Z_{ij}Y_{ij}^2$.
In this way, we bound the following summation $\sum_{j\neq i}\sum_{\substack{k\neq i\\k\neq j}}Y_{ij}(1,1)Y_{ik}(0,0)$ by $\sum_{j\neq i}\Delta (p_{ij}^{-1}Z_{ij}Y_{ij}^2 + q^{-1}_{ik}\overline{Z}_{ik}Y_{ik}^2 )$ for some $\Delta \geq 0$. The additional middle term in \eqref{eq:varest2}, scaled by $\Delta$, implements this conservative correction. 

\begin{theorem}\label{thm:varconserv} 
For Bernoulli randomization with $\min(P_1,P_0)>c$ for some $c>0$, under Assumptions~\ref{asmp:pairitf}--\ref{asmp:bound} and~\ref{asmp:var}, if $\Delta\geq \max_{\mathsf{z}\in\{0,1\}} d_{\infty}(\mathsf{z})$, then
$ {\rm pr}\{\hat{\sigma}^2(\Delta) \geq \sigma^2_{\rm HT}\}\rightarrow 1$ as $n\rightarrow\infty$; that is, $\hat{\sigma}^2(\Delta)$ is asymptotically conservative.
\end{theorem}

Theorem \ref{thm:varconserv} provides a sufficient condition for obtaining a conservative variance estimator.  
The requirement on $\Delta$ depends on the maximum counterfactual degree, which is generally unknown, but it can often be bounded or estimated in practice. 
For instance, suppose interactions can occur only along the edges of a known graph $\mathcal G=(\mathcal V,\mathcal E)$, so that $Y_{ij}(\bm z)\neq\dagger$ only if $(i,j)\in\mathcal E$. In this case, $\max_{\mathsf{z}\in\{0,1\}} d_{\infty}(\mathsf{z})$ is bounded above by the maximum degree of $\mathcal{G}$, and this graph degree can be used as a conservative choice of $\Delta$. 
Alternatively, in the absence of a known graph-based upper bound, one can use a plug-in estimate of the maximum counterfactual degree: 
$\hat d_{\infty} = \max_{1\leq i\leq n}  \sum_{j\neq i} \mathbbm{1}(Y_{ij} \neq \dagger \text{ or }Y_{ji} \neq \dagger) (p^{-1}_{ij}Z_{ij}  + q_{ij}^{-1}\overline{Z}_{ij})$ where $\mathbbm{1}(\cdot)$ is the indicator function.
{This estimator upweights observed treated and control dyads to account for unobserved counterfactual neighbours. By Jensen's inequality, its expectation provides an upper bound on the maximum counterfactual degree. Thus, $\hat d_\infty$ can serve as a practical, conservative choice of $\Delta$ in expectation. The choice of $\Delta$ controls an explicit trade-off: smaller $\Delta$ yields narrower intervals but weaker conservativeness guarantees, while larger $\Delta$ yields wider intervals. When the realized network is approximately degree-regular, $\Delta=d_1$ provides a convenient default; when degree heterogeneity is pronounced, $\Delta=\hat d_\infty$ is a more conservative choice.}

Based on the proposed point and variance estimators, we can construct a Wald-type confidence interval 
$[\hat\tau_{\rm HT} - z_{\alpha/2}\{\hat{\sigma}^2(\Delta)\}^{1/2}, \hat\tau_{\rm HT} + z_{\alpha/2}\{\hat{\sigma}^2(\Delta)\}^{1/2}]$ with nominal coverage $1-\alpha$, where $z_{\alpha/2}$ denotes the upper $\alpha/2$ quantile of the standard normal distribution. 
The null hypothesis $\mathbb H_0:\tau=0$ can be tested using the corresponding Wald statistic. 
In applications, sensitivity analysis over a plausible range of $\Delta$ can be used to assess the robustness of the inferential conclusions. 
{In addition, subsampling provides a useful and computationally attractive approach for handling large networks, and related ideas have been used to obtain approximately independent pairs or subgraphs for valid network inference; see \citet{ogburn2017vaccines} for illustrative applications. }

\section{Simulation study}\label{sec:simu}

\subsection{Simulation with varying sample size, treatment probability, and network degree}\label{sec:simu1}
We conduct numerical simulations to evaluate the performance of the proposed estimators and compare them with competing estimators based on unit-level outcomes. 
We first generate a directed network $\mathcal{G}$ based on a Barab{\\'a}si–Albert scale-free model \citep{albert2002statistical} with average degree $d_1$. 
Then, we generate potential dyadic outcomes with
$Y_{ij}(z_i,z_j) = Y_{ij}(0,0) + \beta_{ij}z_i + \gamma_{ij} z_j + \lambda_{ij} z_{i}z_{j}$, for $z_i,z_j \in \{0,1\}$ for each edge $(i,j)$ in $\mathcal{G}$. 
For each edge $(i,j)$, the parameters are generated independently as $Y_{ij}(0,0)\sim U(0,1)$, $\beta_{ij}\sim U(0,1/2)$, $\gamma_{ij} \sim U(-1/2, 0)$, and $\lambda_{ij}\sim U(0, 1/2)$.
The parameters are fixed across simulation replicates. Treatments are assigned according to a Bernoulli randomization design with $Z_i\sim {\rm Bernoulli}(p)$. 
We consider three simulation scenarios. In the first, we vary the population size over
$n\in\{200,500,1000,1500,2000\}$,
while fixing $p=0.3$ and $d_1=10$. In the second, we vary the treatment probability over
$p\in\{0.1,0.2,0.3,0.4,0.5\}$, while fixing $n=1000$ and $d_1=10$. In the third, we vary the average degree over $d_1\in\{4,10,20,30,40\}$, 
while fixing $n=1000$ and $p=0.3$. Each setting is replicated 500 times.

We compare the proposed estimators with three estimators constructed from unit-level outcomes. The first is the weighted unit-level estimator $\tilde\tau(\bm Y)$; 
the second is the Horvitz-Thompson estimator with an exposure threshold \citep{thiyageswaran2024data}, defined as
\[
\begin{aligned}
    \tilde\tau_{\rm TR}(\bm{Y})=&\frac{\sum_{i=1}^n   \mathbbm{1}(|\mathcal{N}_i|^{-1}\sum_{j\in \mathcal{N}_i} Z_j \geq \lambda_1 ) Z_i Y_i}{\sum_{i = 1}^n  \mathbbm{1}(|\mathcal{N}_i|^{-1}\sum_{j\in \mathcal{N}_i} Z_j  \geq \lambda_1 )Z_i} -\frac{\sum_{i=1}^n   \mathbbm{1}(|\mathcal{N}_i|^{-1}\sum_{j\in \mathcal{N}_i} (1-Z_j) \geq \lambda_0) (1-Z_i) Y_i}{\sum_{i = 1}^n  \mathbbm{1}(|\mathcal{N}_i|^{-1}\sum_{j\in \mathcal{N}_i} (1-Z_j) \geq \lambda_0)(1-Z_i) },
\end{aligned}
\]
for some user-specified threshold $\lambda_1, \lambda_0 \in (0,1)$. We choose $\lambda_1 = p$ and $\lambda_0 = 1-p$. 
This estimator modifies the unit-level estimator by retaining only treated users whose proportion of treated neighbours exceeds $\lambda_1$ and control users whose proportion of control neighbours exceeds $\lambda_0$. 
The last one is the pseudo-inverse estimator \citep{cortez2023exploiting}
\[
\tilde{\tau}_{\rm PI}(\bm{Y}) = \frac{1}{n} \sum_{i=1}^n Y_i \sum_{j \in \mathcal{N}_i}\left(\frac{Z_j}{p_j}-\frac{1-Z_j}{1-p_j}\right).
\]

\begin{figure}[H]
\centering

\subfloat{
    \includegraphics[width=0.54\textwidth]{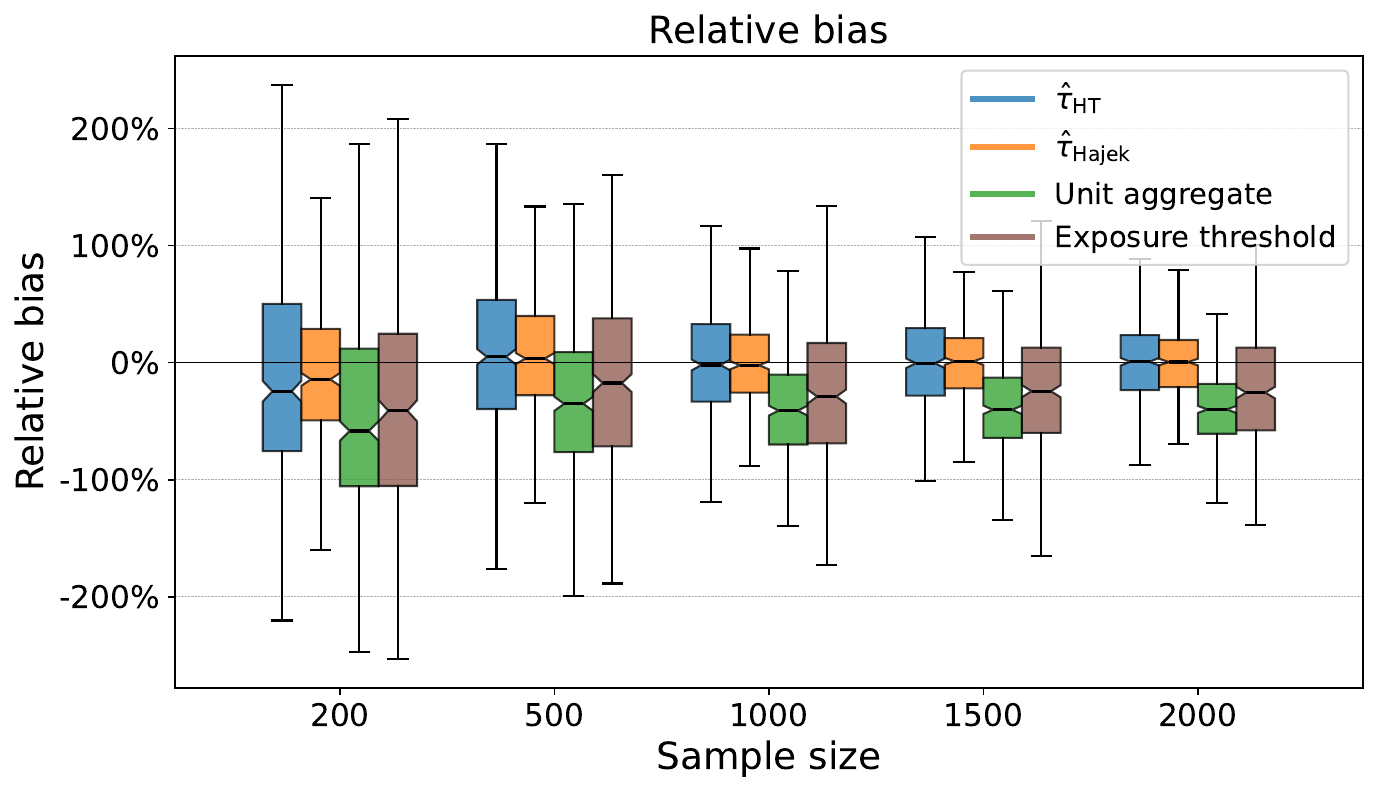}
}
\hfill
\subfloat{
    \includegraphics[width=0.42\textwidth]{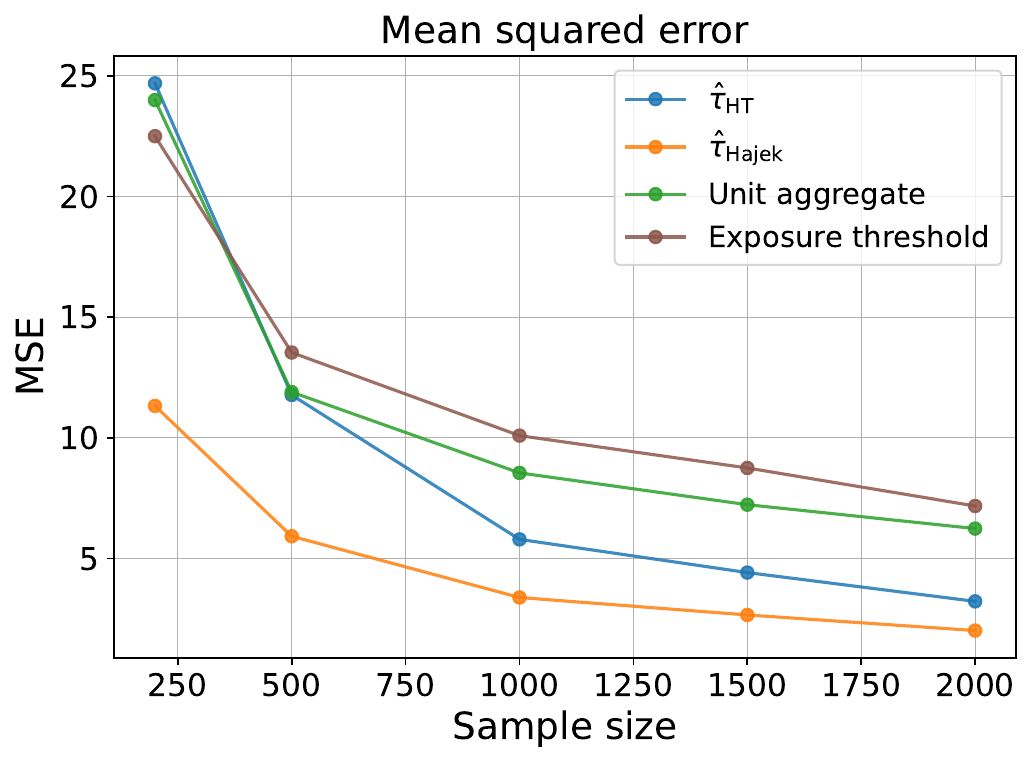}
}

\vspace{4pt}

\subfloat{
    \includegraphics[width=0.54\textwidth]{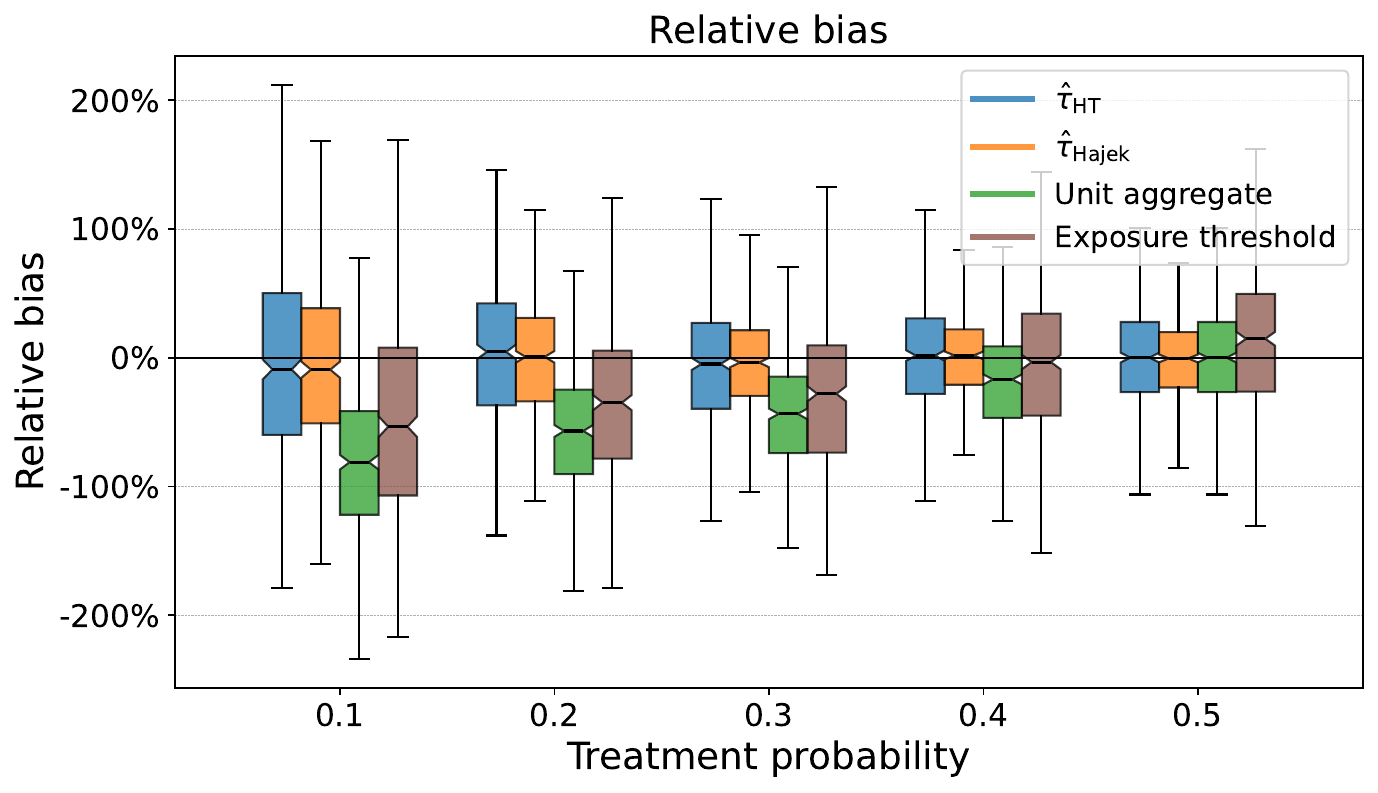}
}
\hfill
\subfloat{
    \includegraphics[width=0.42\textwidth]{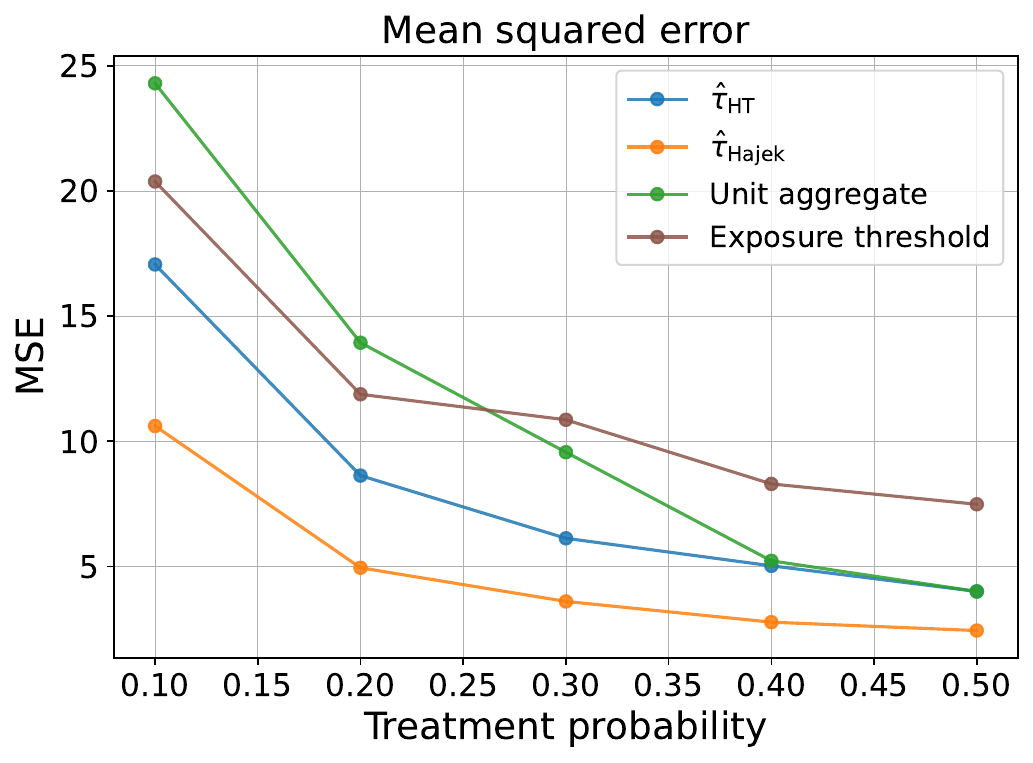}
}

\vspace{4pt}

\subfloat{
    \includegraphics[width=0.54\textwidth]{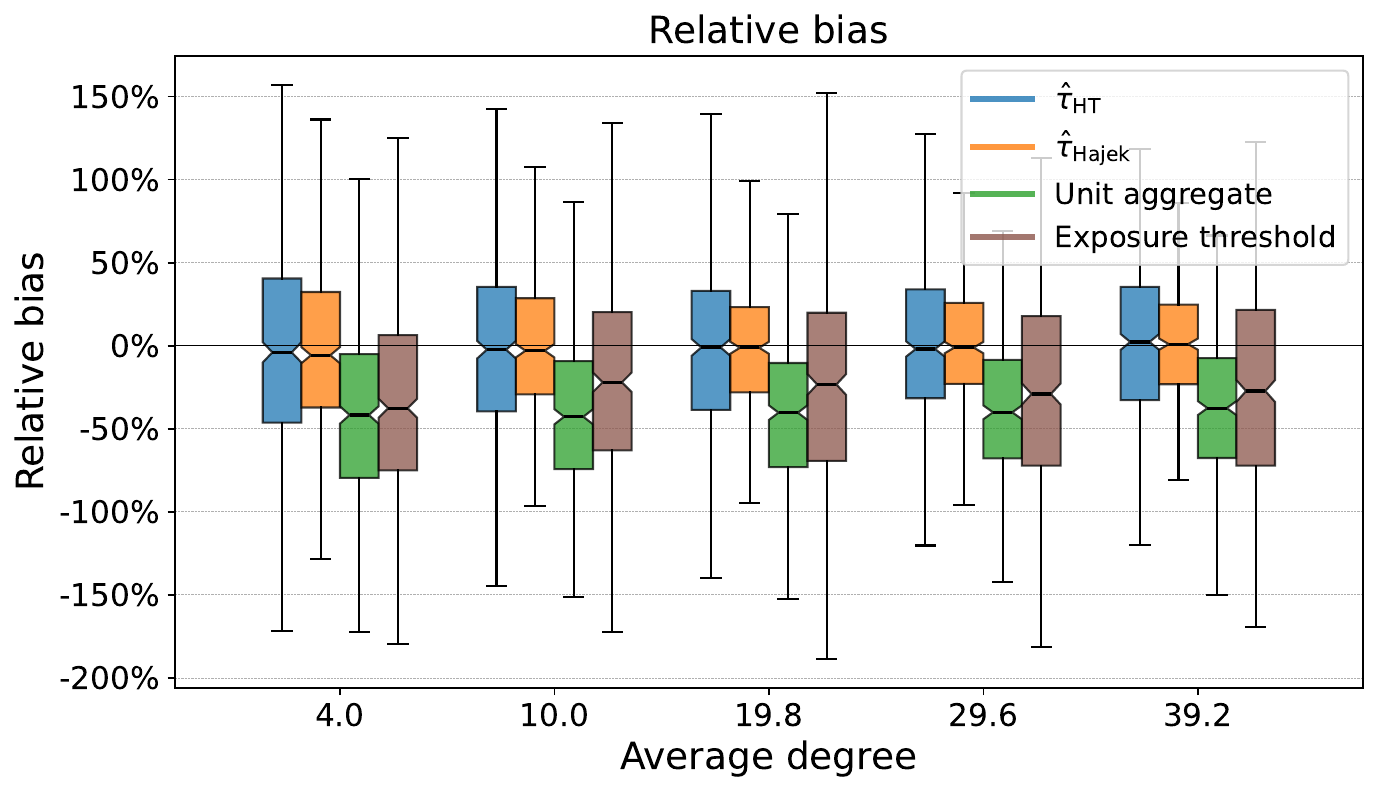}
}
\hfill
\subfloat{
    \includegraphics[width=0.42\textwidth]{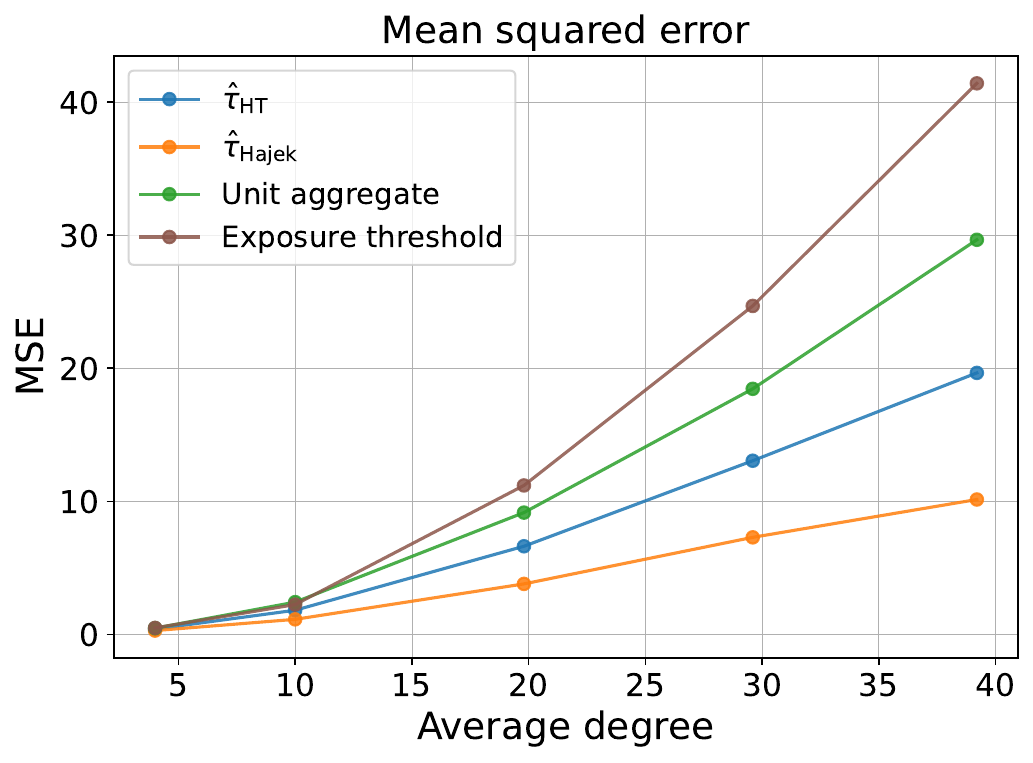}
}

\caption{Relative bias and mean squared error of estimators, across settings with varying sample size, treatment probability, and network degree.}
\label{fig:simu1}
\end{figure}

Figure \ref{fig:simu1} reports boxplots of the relative bias and mean squared error of the estimators across the simulation settings. 
We omit $\tilde\tau_{\rm PI}$ from the figure because its variance is much larger than that of the other estimators in our settings. 
Across these settings, the proposed estimators have small relative bias and lower mean squared error than the unit-level alternatives considered here. The H\'{a}jek estimator is typically more stable than the Horvitz--Thompson estimator in smaller samples, while the difference diminishes as the sample size increases. The unit-level estimators can exhibit  bias because the aggregated outcome does not preserve the dyadic exposure structure.

The simulations also illustrate the roles of network degree and treatment probability. The mean squared error increases as the average degree grows from 2 to 20. When the treatment probability varies, $\tilde\tau(\bm Y)$ is generally biased, although its bias attenuates as $p$ approaches $1/2$. When $p$ is close to zero, the proposed estimators have larger variance due to inverse-probability weighting, but their bias remains much smaller than that of $\tilde\tau(\bm Y)$. These results support the use of $\hat\tau_{\rm H\acute{A}}$ or $\hat\tau_{\rm HT}$ for estimating the global average treatment effect when dyadic outcomes are available.

\begin{figure}[H]
\centering

\subfloat{
    \includegraphics[width=0.315\textwidth]{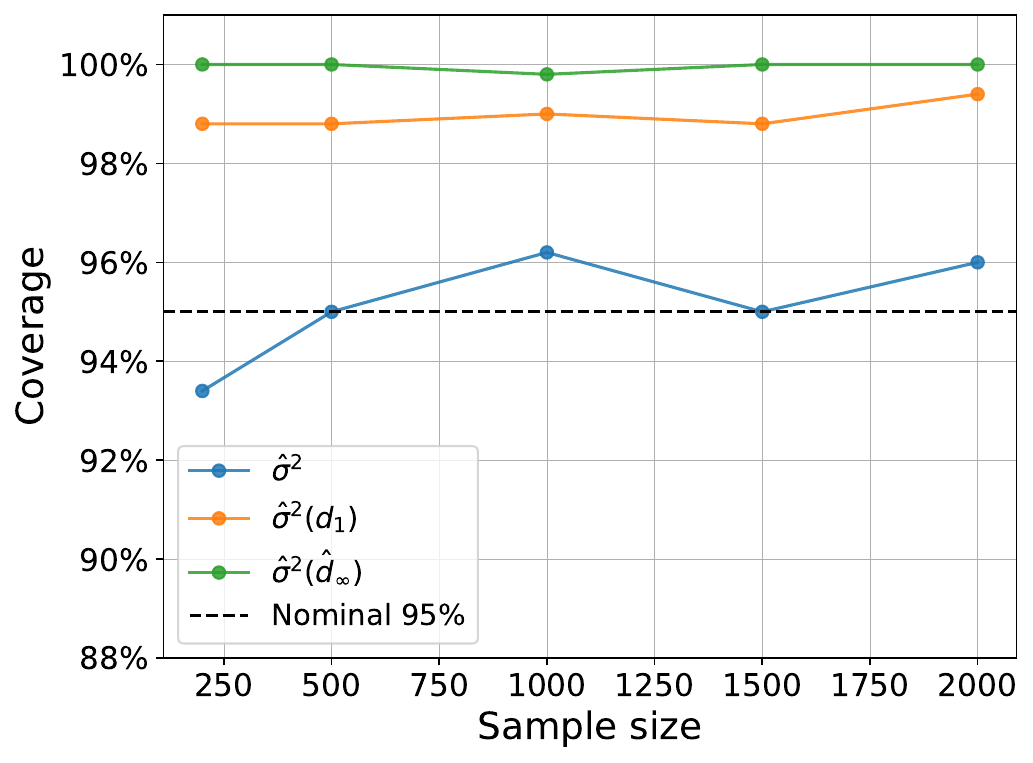}
}
\hfill
\subfloat{
    \includegraphics[width=0.315\textwidth]{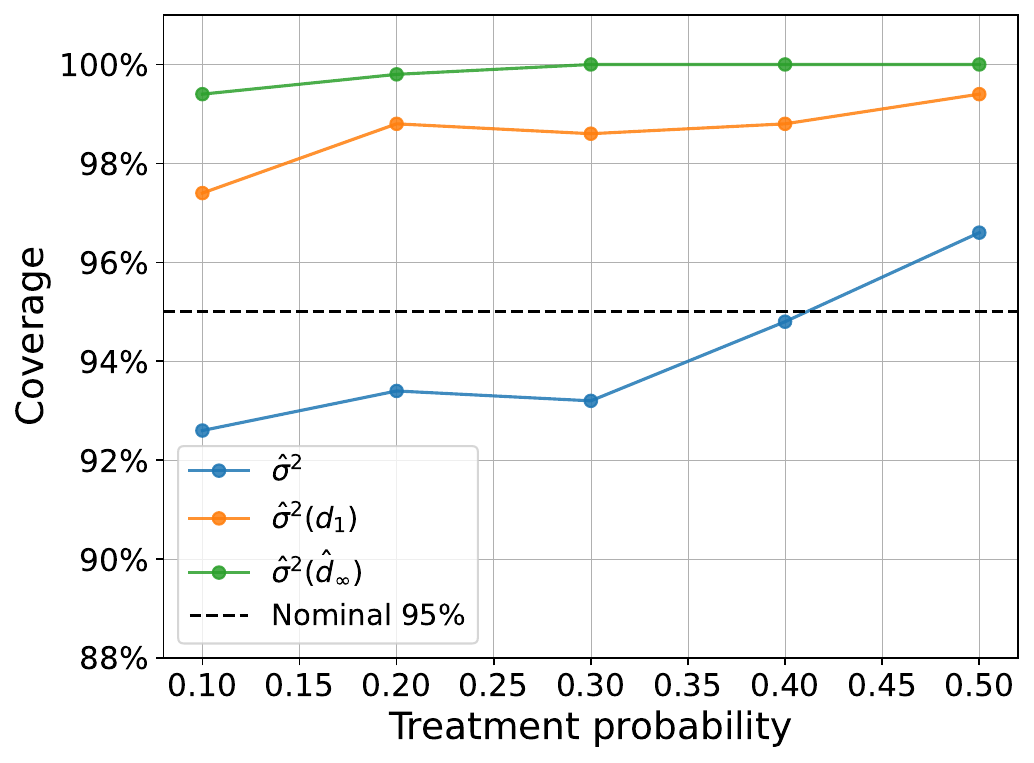}
}
\hfill
\subfloat{
    \includegraphics[width=0.315\textwidth]{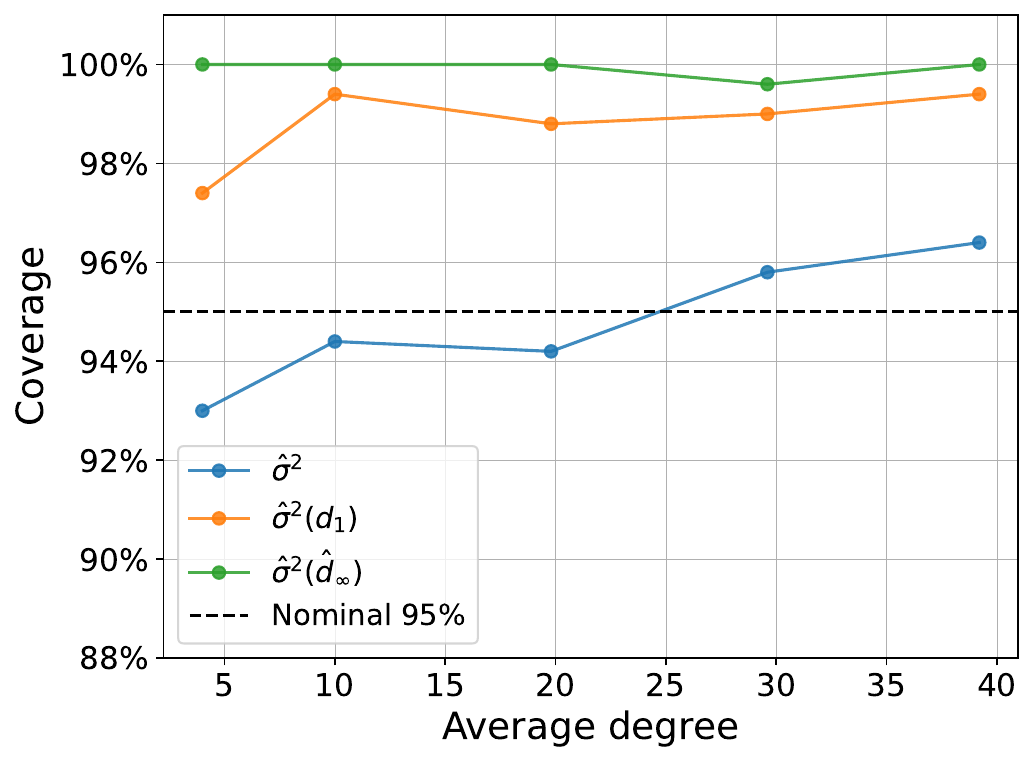}
}

\caption{Coverage rates of 95\% confidence intervals based on different variance estimators, across settings with varying sample size, treatment probability, and network degree.}
\label{fig:simu2}
\end{figure}

\begin{figure}[H]
\centering

\subfloat{
    \includegraphics[width=0.315\textwidth]{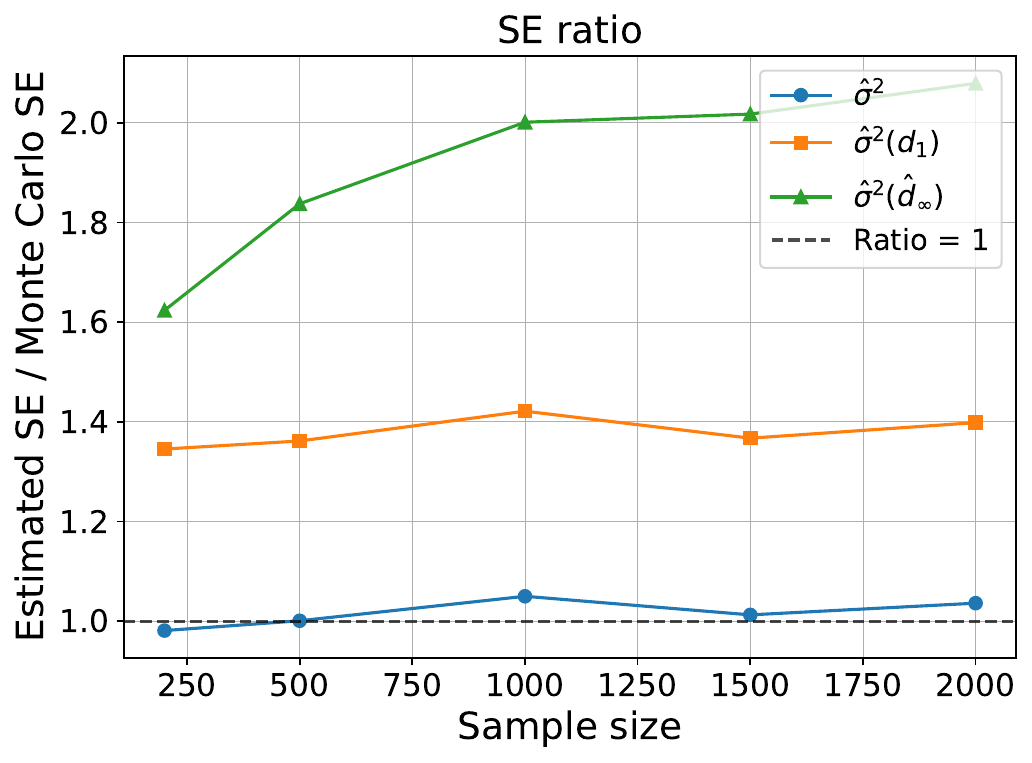}
}
\hfill
\subfloat{
    \includegraphics[width=0.315\textwidth]{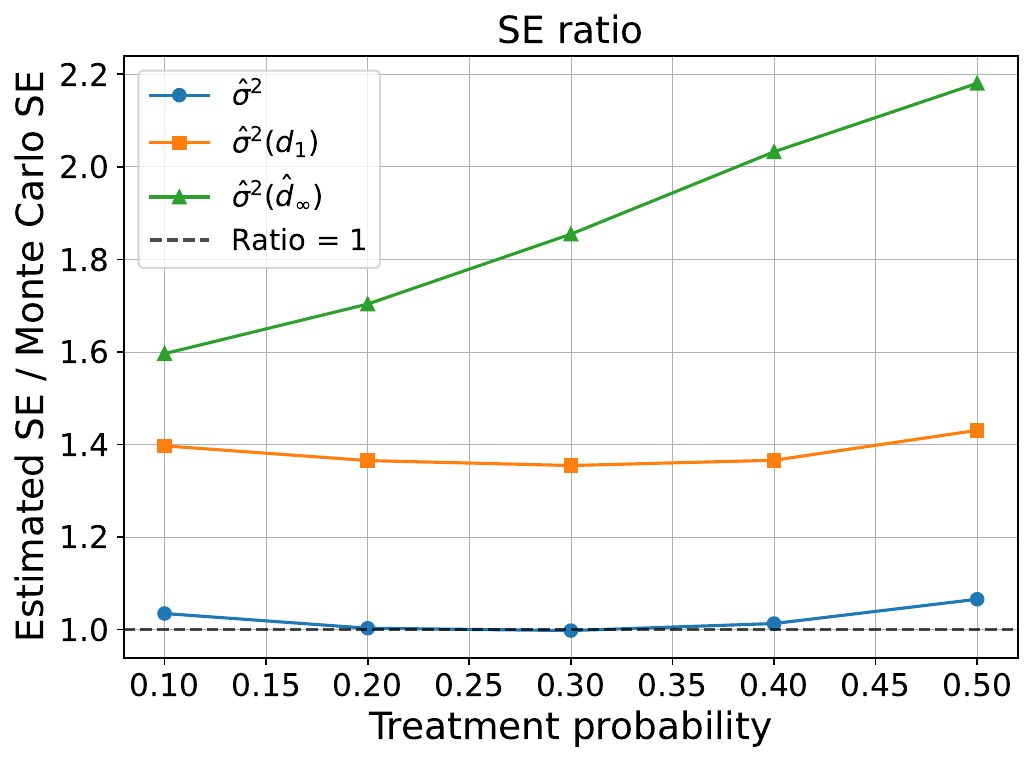}
}
\hfill
\subfloat{
    \includegraphics[width=0.315\textwidth]{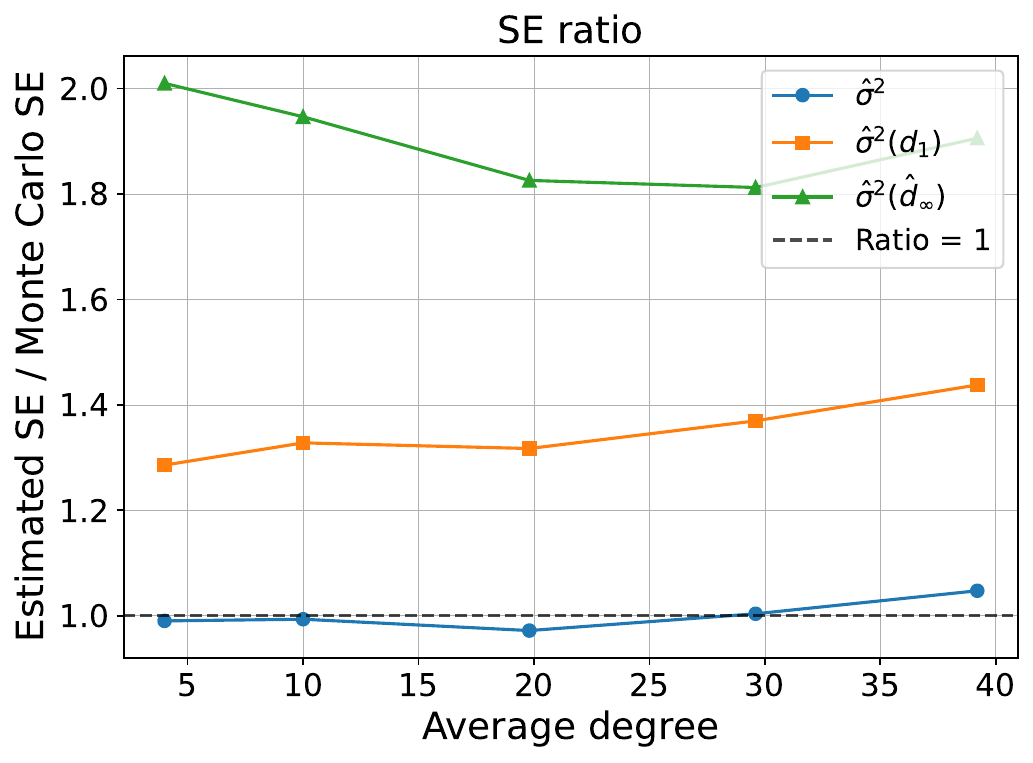}
}

\caption{{Ratio of the average estimated standard error to the Monte Carlo standard error for $\Delta\in\{0,d_1,\hat d_\infty\}$ across varying sample size, treatment probability, and average degree.}}
\label{fig:se_ratio_new}
\end{figure}

To assess the degree of conservativeness, Figure \ref{fig:simu2} summarizes the empirical coverage rate of the Wald-type 95\% confidence intervals based on different variance estimators  with $\Delta\in\{0,d_1,\hat d_\infty\}$. 
Across the simulation settings, intervals based on the unadjusted estimator $\hat\sigma^2$ can undercover, with coverage falling below the nominal level of 0.95. In contrast, intervals based on $\hat\sigma^2(d_1)$ and $\hat\sigma^2(\hat d_\infty)$ achieve coverage above the nominal level, although they can be conservative and sometimes approach full coverage.
{ Figure~\ref{fig:se_ratio_new} reports the ratio between the average estimated standard error and the Monte Carlo standard error. The unadjusted estimator $\hat\sigma^2$ corresponding to $\Delta=0$ often yields a ratio close to one, but it can be insufficient for coverage because it is not guaranteed to be larger than $\sigma^2$. The estimator $\hat\sigma^2(d_1)$ provides a more conservative correction, while $\hat\sigma^2(\hat d_\infty)$ is typically the most conservative, especially when the degree distribution is heterogeneous. These results illustrate the practical trade-off between interval width and robustness: smaller values of $\Delta$ lead to tighter intervals but weaker coverage guarantees, whereas larger values yield wider intervals and higher coverage. We treat $\hat\sigma^2(d_1)$ as an operational degree-adjusted analysis and $\hat\sigma^2(\hat d_\infty)$ as a conservative sensitivity analysis; the latter should therefore be interpreted as robust to substantial degree heterogeneity.
}

\subsection{Performance under  different randomization designs}

We further compare the proposed estimators under Bernoulli and complete randomization. 
We use the Caltech Facebook network \texttt{socfb-Caltech36} \citep{rossi2015network}, which has $n=769$, $d_1 \approx 43.32$, $d_2\approx 3245.95$, and $d_\infty=248$. The data-generating process is the same as that in Section~\ref{sec:simu1}, and we vary the treatment probability over $p\in\{0.1,0.2,0.3,0.4,0.5\}$. 

\begin{figure}[H]
	\centering
		\includegraphics[width=0.97\textwidth]{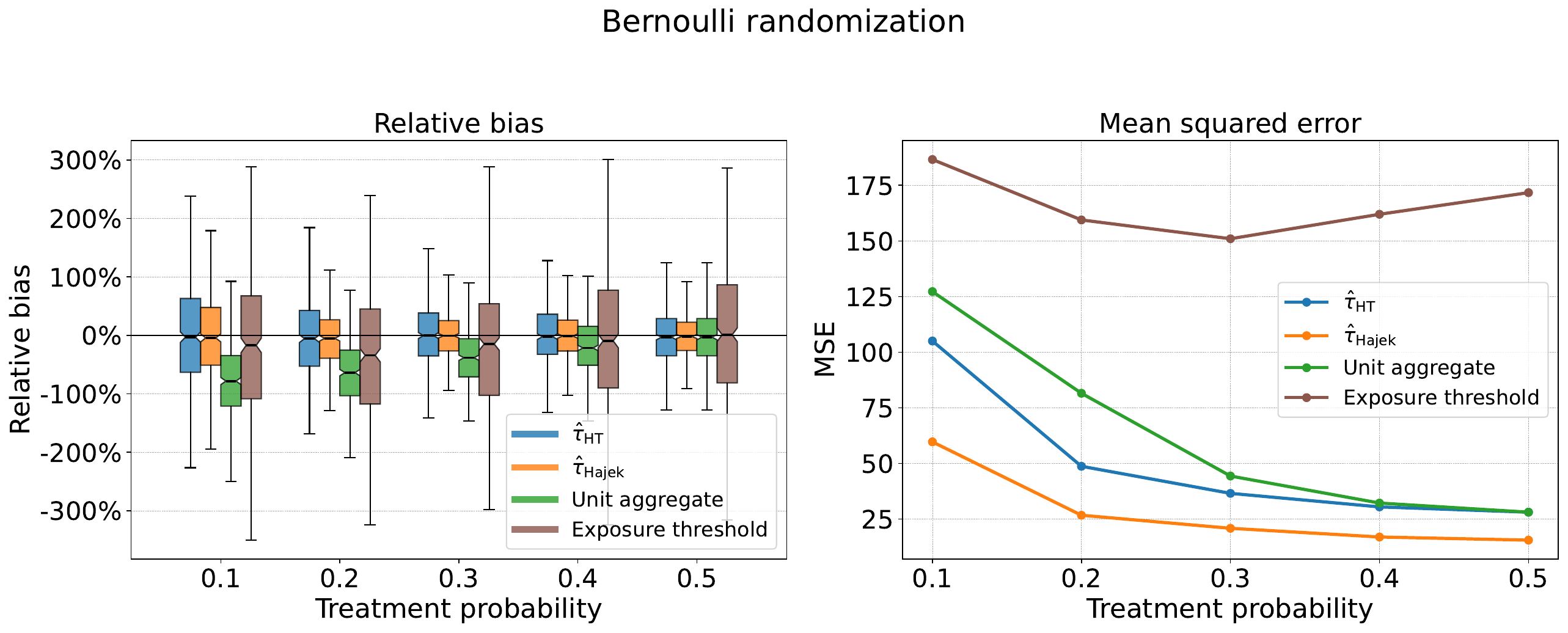}  
            \includegraphics[width=0.97\textwidth]{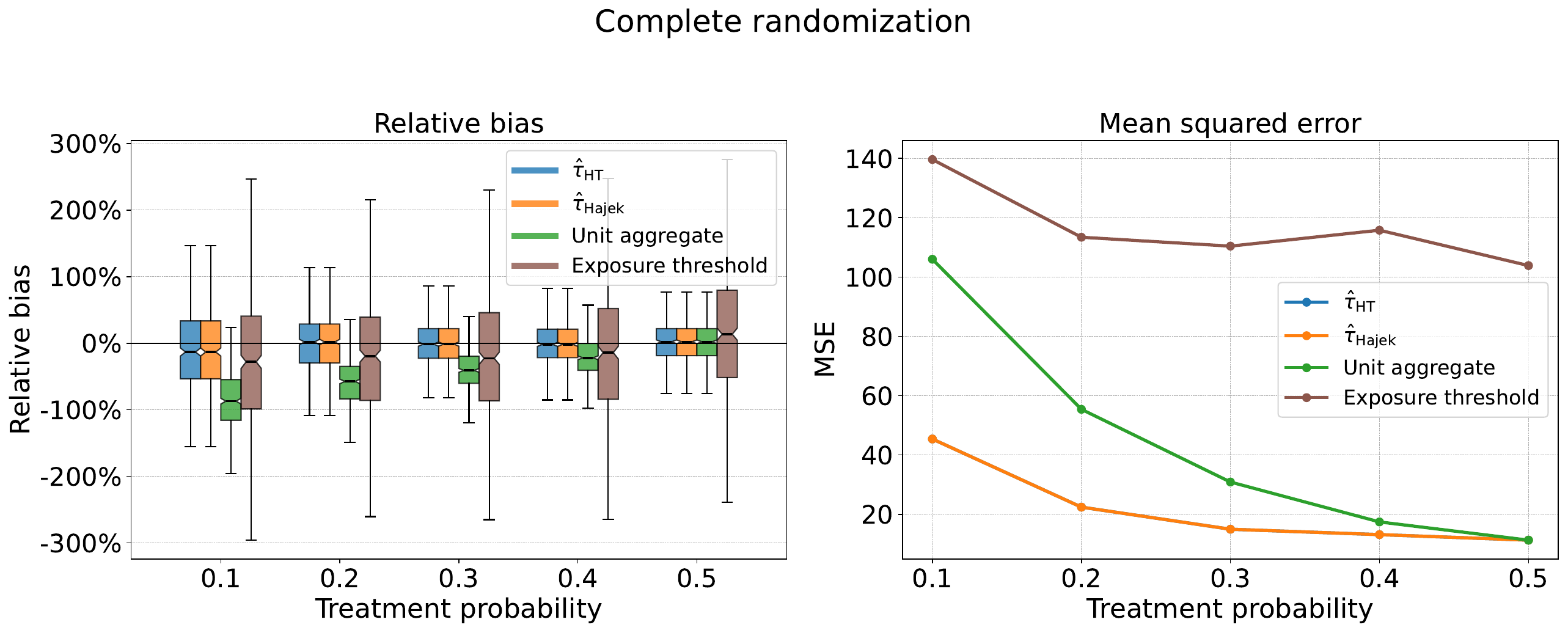}   
	\caption{Relative bias and mean squared error of different estimators under Bernoulli and complete randomization. 
	} \label{fig:simu3}
\end{figure}

Figure \ref{fig:simu3} summarizes the results. For both randomizations, the relative bias and mean squared error of $\hat\tau_{\rm HT}$ and $\hat\tau_{\rm H\acute{A}}$ generally decrease as the treatment probability increases. 
Complete randomization yields smaller mean squared errors than Bernoulli randomization, reflecting the additional stability gained by fixing the number of treated units. As $p$ increases, the performance gap between the two randomization designs becomes less pronounced. Under complete randomization, $\hat\tau_{\rm HT}$ and $\hat\tau_{\rm H\acute{A}}$ coincide, and the resulting estimates are more stable especially when $p$ is small.

{\subsection{Behaviour of the variance estimator under perturbed network degree and propensity}\label{sec:simu_var_behaviour}
We further examine how the variance estimator changes with network degree and treatment probability on the Caltech network. Starting from the baseline graph, we perturb the average degree by randomly adding or deleting edges so that $\Delta d_1\in\{-2,-1,0,1,2,3,5\}$ with $p=0.3$. We also vary the Bernoulli treatment probability over $p\in\{0.1,0.2,0.3,0.4,0.5\}$ while keeping the baseline graph fixed. For each configuration, we generate data as in Section~\ref{sec:simu1}, compute the Monte Carlo variances of $\hat\tau_{\rm HT}$ and $\hat\tau_{\rm H\acute{A}}$ over 1000 replications, and compare them with the average estimated variances $\hat\sigma^2$, $\hat\sigma^2(d_1)$, and $\hat\sigma^2(\hat d_\infty)$. Figure~\ref{fig:var_behaviour_new} shows that the variance increases with network degree and is inflated at low treatment probabilities, in line with the factors involving $d_2(\mathsf z)$ and $P_{\mathsf z}$ in Theorem~\ref{thm:rate}.

\begin{figure}[H]
\centering

\subfloat{
    \includegraphics[width=0.48\textwidth]{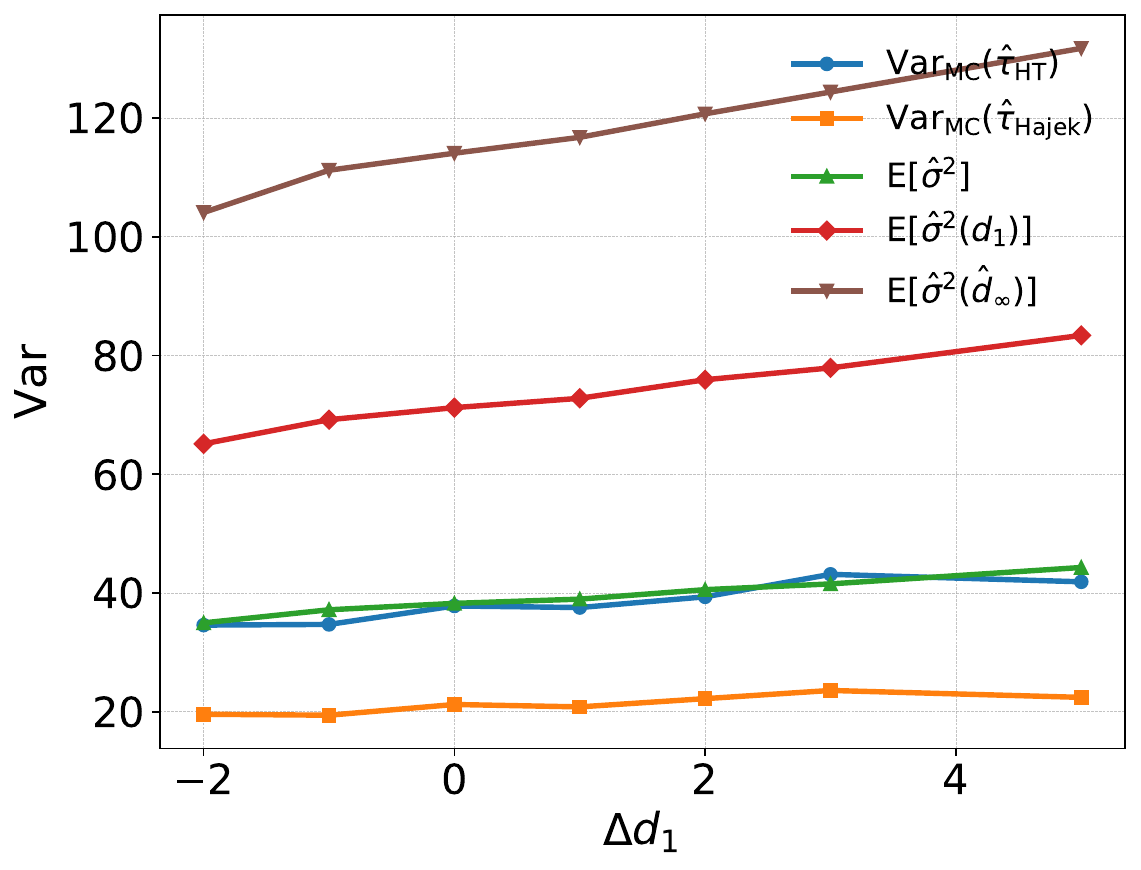}
    \label{fig:var_degree_new}
}
\hfill
\subfloat{
    \includegraphics[width=0.48\textwidth]{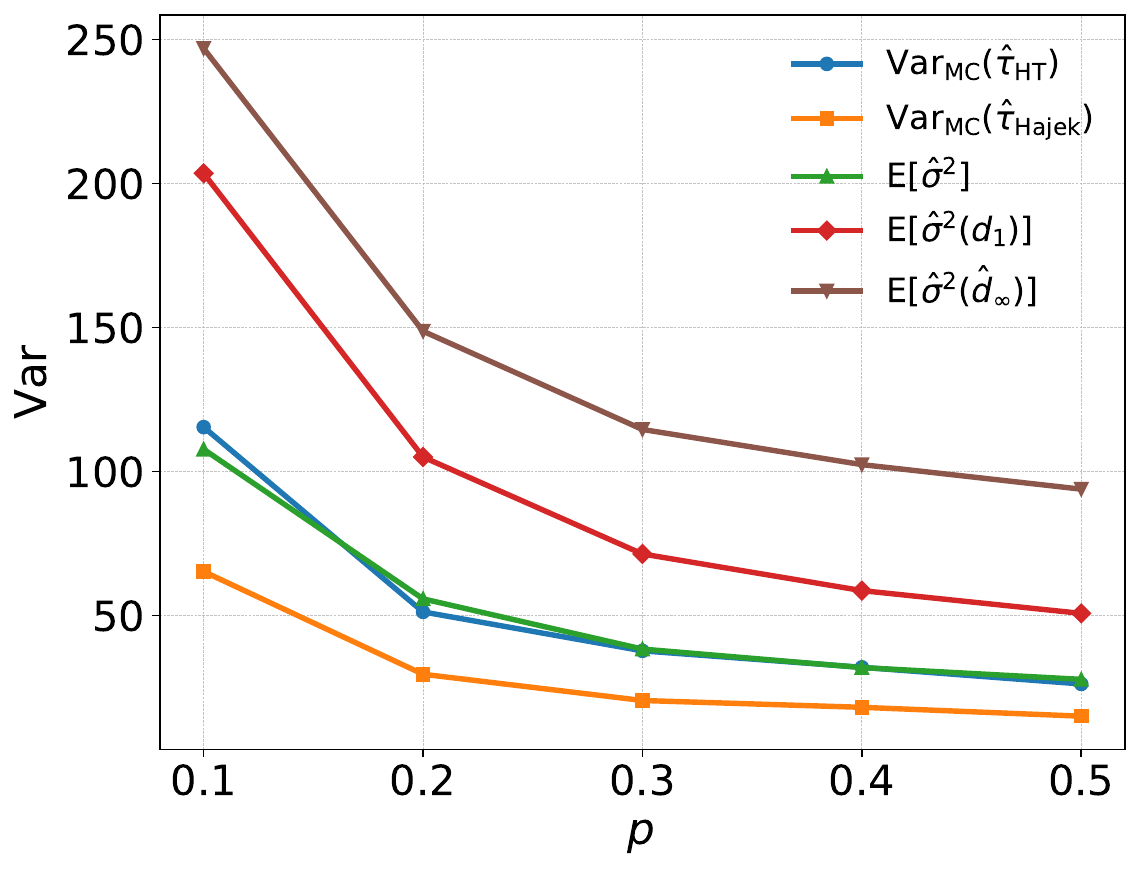}
    \label{fig:var_p_new}
}

\caption{{Monte Carlo variance of $\hat\tau_{\rm HT}$ and $\hat\tau_{\rm H\acute{A}}$ and variance estimators $\hat\sigma^2$, $\hat\sigma^2(d_1)$, and $\hat\sigma^2(\hat d_\infty)$ on the Caltech network.}}
\label{fig:var_behaviour_new}
\end{figure}
Besides these, Section~C of the Supplementary Material studies several subsampling approaches, including dyad-level Poisson subsampling, unit-level Poisson subsampling, and random maximal-matching subsampling for large networks, the latter constructing a maximal set of vertex-disjoint dyads. Although subsampling approaches can improve computational efficiency, our simulation results suggest that they do not alleviate bias from misspecification of dyadic interference, as dyadic outcomes may still depend on treatment statuses outside the sampled dyad.
}

\section{Empirical application}\label{sec:real}

Large-scale technology companies routinely use A/B tests to evaluate product changes, including user-interface updates, recommendation algorithms, and communication-quality improvements. We illustrate the proposed methods using two A/B tests conducted on the WeChat experimentation platform.
The first experiment concerns a content-sharing service within WeChat that allows users to create, share, and discover videos and images beyond their friendship network. WeChat has approximately 1.4 billion monthly active users. In this experiment, treated users received an updated recommendation algorithm designed to increase sharing and clicking activity by ranking content liked by both users and their friends more highly, whereas control users continued to receive recommendations from the previous algorithm. The experiment enrolled approximately $2.4\%$ of all users during the experiment window and used Bernoulli randomization with a treatment probability of $p_i=2/3$.  
{ The second experiment concerns a VoIP call service. The treatment was a new noise-cancellation algorithm that reduces background noise during voice communication. The corresponding dyadic outcomes measure user engagement and experience during VoIP calls, including call-duration-related and user-experience-related metrics. This setting is defined at the dyadic level, because the relevant outcomes are generated through direct interactions between callers and receivers.  The experiment ran for 7 days, enrolled  approximately $10\%$ of all active users, and used Bernoulli randomization with treatment probability $p_i=1/3$. }

We analyse several dyadic outcomes $Y_{m,ij}$ ($m=1,2,3$ for experiment 1 and $m=4,5$ for experiment 2). These metrics capture both user-interaction and system-performance outcomes during the experimental period. The corresponding platform-level metric is $\mathsf{Metric}_{m}=n^{-1}\sum_{i=1}^{n}\sum_{j\in\mathcal{D}_i}Y_{m,ij}$. The targeted estimand is the relative difference
\[
\mathsf{RD}_m =\frac{\sum_{i=1}^{n}\{\sum_{j\in\mathcal{D}_i(\bm{1})}Y_{m,ij}(\bm{1})-\sum_{j\in\mathcal{D}_i(\bm{0})}Y_{m,ij}(\bm{0})\}}{\sum_{i=1}^{n}\sum_{j\in\mathcal{D}_i(\bm{0})}Y_{m,ij}(\bm{0})},
\]
since proportional changes are the standard reporting scale for online experiments. Due to commercial confidentiality, we cannot disclose the exact metric definitions or the absolute sample sizes for the five outcomes. All data were collected with user authorization and were de-identified before analysis. A/A tests and sample-ratio-mismatch diagnostics were conducted to validate the randomization and data-collection pipeline.

{
We estimate the relative difference of the global average treatment effect, $\mathsf{RD}_m$. The dyadic and unit-level  relative-difference estimators are
\[
\widehat{\mathsf{RD}}_{m, {\rm H\acute{A}}}
=\left(\hat{n}_0^{-1}\sum_{i=1}^n \sum_{j\in \mathcal{D}_{i}}q^{-1}_{ij} \overline{Z}_{ij}Y_{m,ij}\right)^{-1}
\hat\tau_{m, {\rm H\acute{A}}},\quad \widetilde{\mathsf{RD}}_m
=\left(n^{-1}\sum_{i=1}^nq^{-1}_{i} \overline{Z}_{i} Y_{m,i}\right)^{-1}
\tilde\tau(\bm{Y}_m).
\]
We test the null hypothesis $\mathbb{H}_0:\tau_m = 0$ for $m=1,\dots,5$. 
The dyadic point estimator is $\hat{\tau}_{m, {\rm H\acute{A}}}$ and the unit-level comparator is $\tilde\tau(\bm{Y}_m)$. We calculate the variance estimators $\hat{\sigma}^2_m$, $\hat{\sigma}^2_m(d_{1})$, and $\hat{\sigma}^2_m(\hat{d}_{\infty})$. 
For each variance estimator, inference is based on normal approximation, and the two-sided $p$-value is computed as $2[1-\Phi\{|\hat\tau_{m}|/\hat{\sigma}_m(\Delta)\}]$, where $\Phi(\cdot)$ denotes the standard normal CDF.}

\begin{table}[H]
\centering
\caption{Relative difference, standard errors, p-values, and estimated sample sizes for the estimators. {Standard errors are expressed on the relative-difference scale by normalizing the original standard errors with the estimated mean of all-control dyads. The quantity $n^*/n$ is the sample-size multiplier required to achieve a p-value below $0.05$ under $\hat{\sigma}_m(\hat d_\infty)$.}}
{
\setlength{\tabcolsep}{13pt}
\renewcommand{\arraystretch}{0.9}
\begin{tabular}{@{}lccccc@{}}
\toprule
\multirow{2}{*}{Quantity} 
& \multicolumn{3}{c}{Experiment 1} 
& \multicolumn{2}{c}{{Experiment 2}} \\
\cmidrule(lr){2-4}\cmidrule(lr){5-6}
& $m=1$ & $m=2$ & $m=3$ 
& {$m=4$} 
& {$m=5$} \\
\midrule

$\widehat{\mathsf{RD}}_{m,{\rm H\acute{A}}}$ (\%) 
& 1.012 & 0.624 & 1.124 
& {0.072} & {0.061} \\

{SE $[\hat{\sigma}_m]$} (\%)          
& 0.776 & 0.307 & 0.498 
& {0.024} & {0.021} \\

{SE $[\hat{\sigma}_m(d_1)]$}(\%)     
& 0.842 & 0.333 & 0.587 
& {0.033} & {0.031} \\

{SE $[\hat{\sigma}_m(\hat d_\infty)]$} (\%) 
& 1.788 & 0.712 & 0.968 
& {0.046} & {0.044} \\

p-value $[\hat{\sigma}_m]$           
& 0.193 & 0.042 & 0.024 
& {0.003} & {0.004} \\

p-value $[\hat{\sigma}_m(d_{1})]$    
& 0.230 & 0.061 & 0.055 
& {0.029} & {0.049} \\

p-value $[\hat{\sigma}_m(\hat{d}_{\infty})]$ 
& 0.571 & 0.381 & 0.245 
& {0.118} & {0.166} \\

{$n^*/n$}                             
& 12.0 & 5.0 & 2.9 
& {1.6} & {2.0} \\

\midrule

$\widetilde{\mathsf{RD}}_m$ (\%)        
& 0.076 & -0.120 & 0.285 
& {0.018} & {0.026} \\

SE (\%)  
& 0.336 & 0.327 & 0.497 
& {0.016} & {0.015} \\

p-value                                
& 0.821 & 0.713 & 0.566 
& {0.261} & {0.083} \\

\bottomrule
\end{tabular}
}
\label{tab:ace}
\end{table}

{

Table~\ref{tab:ace} reports estimated relative differences and p-values for the dyadic H\'{a}jek estimator and the unit-level estimator. In Experiment~1, the unit-level estimator gives relative differences close to zero for $m=1,2$ and a small positive value for $m=3$, none of which has p-values below 0.05. The dyadic estimator gives larger positive estimates for all three metrics. The effects for $m=2$ and $m=3$ are significant under the unadjusted variance estimator, whereas the degree-adjusted estimators yield larger standard errors and weaker evidence, illustrating the coverage--power trade-off discussed in Section~\ref{sec:confidence}.
In Experiment~2, the proposed estimator gives positive relative differences for both VoIP-related metrics. Both are significant under the unadjusted variance estimator and remain significant under $\hat\sigma_m^2(d_1)$, while the evidence weakens under the most conservative estimator $\hat\sigma_m^2(\hat d_\infty)$. Taken together, the two experiments suggest that dyadic analyses can reveal pairwise interaction effects that are attenuated after aggregation to unit-level outcomes, while also making explicit the inferential cost of conservative variance estimation.}

\section{Discussion}\label{sec:dis}

This paper develops a design-based framework for randomized experiments in which the outcomes of interest are dyadic interactions. By analyzing the dyadic outcomes directly, the proposed estimators target the global average treatment effect. The theory shows how the properties of estimators depend on counterfactual network degrees, making explicit the role of network sparsity and degree imbalance.

The framework can be extended to settings in which dyadic and unit-specific outcomes are both of substantive interest. In the link-sharing example, $Y_{ij}$ may denote the number of videos shared by user $i$ and subsequently viewed by user $j$, while $Y_{ii}$ may capture videos viewed by user $i$ through channels that are not attributable to a specific sharing dyad, such as search or recommendation feeds. In this case, the upstream aggregate can be defined as
$U_i=Y_{ii}+\sum_{j\in\mathcal U_i}Y_{ji}.$
When $Y_{ii}$ depends only on the unit's own treatment assignment, i.e., $Y_{ii}(\bm z)=Y_{ii}(z_i)$, the Horvitz--Thompson estimator is modified by including an additional unit-level term
$
\hat{\tau}_{\operatorname{Mix-HT}} = \hat{\tau}_{\rm HT}
 + n^{-1}\sum_{i=1}^n\left\{p_i^{-1}Z_i-q_i^{-1}(1-Z_i)\right\}Y_{ii},
$
and its large-sample properties follow by combining the dyadic arguments with standard design-based arguments for unit-level outcomes.

Several extensions merit further study. First, pre-experiment covariates may be used in the design stage through stratified randomization, matched-pair designs, or covariate-adaptive assignment. In observational studies, the propensity scores are unknown and must be estimated. Extending the present framework to incorporate covariates and estimated propensities could improve efficiency while preserving robustness to dyadic interference. {Second, the main limitation is that the dyadic interference Assumption~\ref{asmp:pairitf} rules out higher-order propagation of treatment effects across paths of length greater than one.} A sensitivity analysis can be formulated by allowing
\[
Y_{ij}(\bm z)=\alpha_{ij}+\beta_{ij}z_i+\gamma_{ij}z_j+\lambda_{ij}z_iz_j+\zeta_{ij}\frac{\sum_{k\in\mathcal U_i(\bm z)}z_k}{|\mathcal U_i(\bm z)|},
\]
where $\zeta = \max_{ij}|\zeta_{ij}|$ bounds the magnitude of indirect upstream effects through the treated fraction of upstream neighbours. Varying $\zeta$ over a plausible range can quantify the sensitivity of the estimators to violations of dyadic interference, see \citet{vanderweele2015interference} for the sensitivity analysis with interference. In the Supplementary Material, we provide simulation evidence illustrating how such misspecification affects bias and mean squared error under increasing levels of higher-order spillovers. Developing more general sensitivity procedures remains an important direction for future work.

\bibliographystyle{agsm}
\bibliography{reference}

\end{document}